\documentclass[12pt]{article}
\usepackage{amsfonts,amsmath}
\usepackage{amssymb}
\usepackage[hypertex]{hyperref}

\makeatletter \@addtoreset{equation}{section} \makeatother

\newcommand{\hhmt}{{{h}}}

\newcommand{\hmt}{{\vartriangle}}%{{\rm{d}^*}}%{{\mathfrak{h}}}% {{  {\Delta^{-1}}{}}}%
\newcommand{\be}{\begin{equation}}
\newcommand{\ee}{\end{equation}}
\newcommand{\bee}{\begin{eqnarray}}
\newcommand{\beee}{\begin{array}}
\newcommand{\eee}{\end{eqnarray}}
\newcommand{\eeee}{\end{array}}

\newcommand{\De}{\mathbb{D}}
\newcommand{\ga}{\alpha}

\newcommand{\gb}{\beta}
\newcommand{\gga}{\gamma}

\newcommand{\M}{{\cal M}}

\newcommand{\E}{{\cal E}}

\newcommand{\W}{{\cal W}}

\newcommand{\rhs}{{\it r.h.s.} }
\newcommand{\rhss}{{\it r.h.s.} }
\newcommand{\lhs}{{\it l.h.s.} }

\newcommand{\ie}{{\it i.e.,} }
\newcommand{\ls}{\!\!\!\!\!\!}

\newcommand{\gep}{\epsilon}

\newcommand{\gs}{\sigma}

\newcommand{\go}{\omega}

\newcommand{\by}{{\bar{y}}}

\newcommand{\q}{\,,\qquad}

\newcommand{\dga}{{\dot{\alpha}}}
\newcommand{\dgb}{{\dot{\beta}}}

\newcommand{\nn}{\nonumber}

\newcommand{\p}{\partial}

\newcommand{\D}{{\cal D}}

\newcommand{\f}{\frac}
\newcommand{\ff}{\frac}
\newcommand{\B}{{\cal B}}%%

\newcommand{\V}{{\cal V}}

\newcommand{\g}{\gamma}
\newcommand{\Go}{{\Omega}}
\newcommand{\Ex}{{\E(\Omega)}}

\newcommand{\Exp}{{\E(\Omega')}}
\newcommand{\Exb}{{\E(\bar \Omega)}}
\newcommand{\dal}{\dot \alpha}

\newcommand{\al}{ \alpha}
\newcommand{\dr}{{\rm d}}

\begin{document}

\begin{flushright}
%\texttt{Started on May 14}
%\\
{\small FIAN/TD/13-23}
\end{flushright}
\vspace{1.7 cm}

\begin{center}
{\large\bf Differential Contracting Homotopy  in  Higher-Spin
Theory}

\vspace{1 cm}

{\bf  M.A.~Vasiliev}\\
\vspace{0.5 cm}
{\it
 I.E. Tamm Department of Theoretical Physics, Lebedev Physical Institute,\\
Leninsky prospect 53, 119991, Moscow, Russia}

\end{center}

\vspace{1.2cm}

{\it $\phantom{MMMMMMMMMMMMMMMMMM}$ To the memory of Lars Brink}

\vspace{0.8 cm}

\begin{abstract}
A new efficient approach to the analysis of nonlinear higher-spin equations, that
treats democratically auxiliary spinor variables $Z_A$ and integration homotopy parameters
 in the non-linear vertices of the higher-spin theory,  is developed. Being most general,
 the proposed approach is the same time far simpler than those available  so far.
 In particular, it is free from the necessity to use the Schouten identity. Remarkably,
 the problem of reconstruction of higher-spin vertices is mapped  to certain  polyhedra cohomology in terms of homotopy parameters themselves. The new scheme provides a
 powerful tool for the study of higher-order corrections in higher-spin theory and,
 in particular, its spin-locality. It is illustrated by the analysis of the lower
 order vertices, reproducing not only the results  obtained  previously  by the
 shifted homotopy approach but also projectively-compact vertices with the minimal
 number of derivatives, that were so far unreachable within that scheme.

\end{abstract}

\newpage
%\textheight 22.6 true cm
\tableofcontents
\newpage

\section{Introduction}

\subsection{General background}

Higher-spin (HS) gauge theory is a nonlinear theory of HS massless fields of spins $s>2$
originally identified as gauge fields by Fronsdal \cite{Fronsdal:1978rb} at the free
field level. At the nonlinear level, HS gauge theory
 admits a natural formulation in the Anti-de Sitter space
\cite{Fradkin:1987ks}. The HS gauge theory is organized by the symmetry
principle resulting from gauging
the infinite-dimensional HS algebra found in \cite{Fradkin:1986ka}.

That HS symmetry is infinite dimensional has several important consequences.
First of all this implies that HS gauge theory describes infinite towers of
fields of unlimited spins as was first noticed in \cite{Berends:1984rq}.\footnote{We discuss HS theories in $d\geq 4$
with propagating HS fields. In $d<4$ HS theories with topological HS fields
this is not necessarily the case (see, e.g., \cite{Vasiliev:1989qh}-\cite{Prokushkin:1998bq}).} By construction, HS symmetries are extensions
of the conventional space-time symmetries (usually in $(A)dS_d$) with the generators $t_i$.
The full set of generators $T_A$ contains space-time generators $t_i$ and HS generators
$T_A^{HS}$
\be
T_A = (t_i, T_A^{HS})\,.
\ee
The generators $t_i$ form a space-time algebra $s$,
\be
[t_i\,,t_j] = f_{ij}^k t_k.
\ee
The commutation relations between space-time and HS generators have the structure
\be
\label{tT}
[t_i, T^{HS}_A ] = f_{iA}^B T_B^{HS}\,.
\ee
Since in Cartan formulation of gravity, the gauge fields associated with
$t_i$ describe a spin two vielbein and Lorentz connection, (\ref{tT})
implies that HS symmetry transforms the graviton to a
HS field. Hence, the metric tensor is not an invariant concept in presence
of HS symmetries. Since the metric tensor is a tool for measuring a distance
between space-time points, this elementary observation suggests that the issue of
locality may be  nontrivial in the HS gauge theory. Practically, this is manifested
as follows.
Being consistently formulated in $AdS$ background \cite{Fradkin:1987ks}, HS gauge theories contain
higher derivatives  of degrees
increasing with spins in interactions \cite{Bengtsson:1983pd,Berends:1984rq,Fradkin:1987ks}, and involve infinite towers of fields of unlimited
spins \cite{Berends:1984rq,Fradkin:1986ka}. The fact that HS cubic vertices
contain higher derivatives with the number of derivatives increasing with spin
 was originally found in the light-front formalism in the seminal paper by
Anders Bengtsson, Ingemar Bengtsson, and Lars Brink \cite{Bengtsson:1983pd}.
It should be stressed that any vertex for a finite subset of spins is local,
containing a finite number of derivatives depending on the spins in the vertex,
which property is called {\it spin-locality} \cite{Gelfond:2018vmi}.

Being one of the founders of String Theory, Lars Brink
influenced a lot HS theory as well. Having a lucky opportunity to
discuss  fundamental physics with Lars in person, I can confirm that
 both of us shared an opinion that String Theory is closely related  to HS
theory and, most likely,  is a spontaneously broken realization
 of the latter.  Though the realization of this program
(see, e.g., \cite{Vasiliev:1987zv}-\cite{Gross:1988ue}) is still
incomplete, a better understanding of the underlying symmetries was reached
recently in \cite{Vasiliev:2018zer}, that gives a hope to find a solution to
this fundamental problem relatively soon. It is worth noted that the construction of
\cite{Vasiliev:2018zer} is based on the algebra of observables of the $N$-body Calogero
model, introduced in collaboration with Lars Brink, Hans Hansson and Semyon Konstein in
\cite{Brink:1992xr,Brink:1993sz}, and its further multiparticle extension
\cite{Vasiliev:2018zer} to the Coxeter group $B_2$, while the
simplest HS algebra \cite{Vasiliev:1989qh,Vasiliev:1989re} corresponds to the
two-body Calogero case of $A_1=\mathbb{Z}_2$.

Another fundamental problem in HS theory also closely related to the studies of Lars
and collaborators is to understand the
degree of non-locality of HS theory  still debatable in the literature within
various formalisms. One of the most popular and seemingly simplest ones is
based on the holographic correspondence
\cite{Maldacena:1997re}-\cite{Witten:1998qj}.
Klebanov-Polyakov conjecture suggested that HS theory is holographically dual
 to a $3d$ vector boundary sigma model \cite{KP}.
(See also \cite{Sezgin:2002rt}; further generalizations were worked out to supersymmetric
\cite{LP,Sezgin:2003pt}
and Chern-Simons extensions    \cite{Aharony:2011jz,Giombi:2011kc}.)
Though Klebanov-Polyakov conjecture
  obeys  kinematic constraints of the linearized holography
expressed by the Flato-Fronsdal theorem \cite{FF} implying the relation between
currents (tensor products) built from free $3d$ conformal  fields
and free massless fields in $AdS_4$, attempts to
reconstruct HS interactions in the bulk led the authors of
\cite{Bekaert:2015tva}-\cite{Neiman:2023orj}
 to the conclusion
that HS gauge theory must be essentially non-local beyond the leading order.

Alternatively, HS gauge theory can be studied directly in the bulk.
Apart from the standard Noether procedure \cite{Berends:1984rq}  and
its further extension by the BV-BRST formalism (see e.g.
\cite{Bekaert:2005jf}-\cite{Buchbinder:2023xlb})
typically efficient at the lower orders, there are two
important tools to simplify the problem.
One is the light-cone formalism initiated by Lars Brink and
collaborators in \cite{Bengtsson:1983pd} and
further developed by A.~Bengtsson  \cite{Bengtsson:2012jm}, Metsaev
\cite{Metsaev:1991mt}-\cite{Metsaev:2007rn},
 and others (see e.g. \cite{Ivanovskiy:2023aay}).
 Since HS holography is weak-weak \cite{KP,Giombi:2012ms} it is important to study
HS gauge theory  both holographically and in the bulk independently.

An efficient covariant approach of
\cite{Vasiliev:1992av}  makes it possible  to
reconstruct on-shell HS vertices order by order in the so-called unfolded formalism.
This way, using shifted homotopy approach, in
\cite{Gelfond:2018vmi}, \cite{Didenko:2018fgx}-\cite{Gelfond:2021two}
some higher-order vertices
in HS theory were reconstructed  that all have been shown to be spin-local.
(For the closely related though somewhat different approaches for the reconstruction of the so-called holomorphic HS
vertices see also \cite{Didenko:2022qga,Sharapov:2022nps}.\footnote{Both of these approaches
use the same limiting HS algebra as in \cite{Didenko:2019xzz}, facing however potential
divergency problems  due to the lack of the well-defined limiting procedure of
\cite{Didenko:2019xzz}, that are argued not to contribute to the final result in
\cite{Didenko:2022qga} or  being wishfully regularizable in \cite{Sharapov:2022nps}. })
In \cite{Vasiliev:2022med} a sufficient condition was formulated that guarantees
equivalence of space-time and spinor (twistor) spin-locality, requiring the vertices to
be not only spin-local in the spinor space but also "projectively-compact". It was also noted
in \cite{Vasiliev:2022med} that the vertices found earlier in \cite{Vasiliev:2016xui,Gelfond:2017wrh}
belong to the projectively-compact class while those found in \cite{Didenko:2020bxd} do not.
The difference between the two types of vertices is spin-local:
vertices of \cite{Vasiliev:2016xui,Gelfond:2017wrh} contain the minimal possible number of
space-time derivatives while those of \cite{Didenko:2019xzz} contain local improvement terms.
Naively, that makes no difference. That would  indeed be true in the models with a
finite number of fields but may not be true in the HS theories due to infinite summations over
spins \cite{Vasiliev:2022med}.
So far it was not clear how to reproduce the vertices of \cite{Vasiliev:2016xui,Gelfond:2017wrh} within
the shifted homotopy formalism of \cite{Didenko:2020bxd}  without  a
by hand local field redefinition.

The formulation of \cite{Vasiliev:1992av} is complete in the sense that
it represents any solution to the problem including all possible field
redefinitions. This has both advantages and disadvantages. The disadvantage
is that equations of \cite{Vasiliev:1992av} do not directly lead to a local
or minimally non-local form of the equations whatever it is. In other words, to proceed
one has to work out appropriate additional conditions  that single out
a proper form of HS field equations. This is somewhat analogous
to the Schroedinger equation in QM: most of its
solutions  have no relation to physics.
One has to, first, impose an additional condition that the wave-function must
belong to $L_2$ and then find  appropriate solutions.
Analogously, in the analysis of HS gauge theory based on the equations of
\cite{Vasiliev:1992av} a list of conditions has been identified in
 \cite{Gelfond:2018vmi} that reduce the degree of non-locality of the
resulting vertices. As a result, it was
shown that all lowest-order
vertices are properly reproduced by the equations of \cite{Vasiliev:1992av}.
These results were  then extended \cite{Didenko:2020bxd} to a class of
non-linear vertices of the types discussed in \cite{Sleight:2017pcz}.

Within  the shifted homotopy technique,
different versions of the HS theory associated with different choices of field variables
result from different shifted homotopy procedures that depend on the homotopy parameters
$\tau$ and so-called shift parameters $ \rho,\gb,\gs, \ldots$.
Being  efficient in the
lowest orders, the original shifted homotopy approach is less efficient at higher orders.
In particular, the higher-order locality condition seemingly suggests \cite{GV} that higher-order shift
homotopy parameters must depend on those that appeared at the previous orders. Naively, this is
impossible since the homotopy parameters that appeared at the lower orders have been already
integrated. Nevertheless, as shown in this paper there exists an extension of the formalism
in which the homotopy and shift parameters can be treated as local coordinates in a
larger space $\M$, that extends space-time $M$, while the HS vertices are understood as
integrals of the differential forms in $\M$. (Note that $\dim \M$ increases with
the order of the vertex and it is  usually convenient to realize $\M$ as a polyhedron
rather than a smooth manifold.)
The proposed formalism is invariant under diffeomorphisms in $\M$ that gives a simple explanation to some of the
obscure relations in the previous approach. Most remarkably, however, it trivializes
the role of the Schouten identities in the theory that was a source of headache in the standard formalism
like that of \cite{Gelfond:2021two}. Let us mention that the idea of the approach of this paper has much in common
with the approach of Gelfond and Korybut \cite{Gelfond:2021two} where HS vertices were reconstructed without direct
use of the homotopy technique.
Numerous nice properties of the proposed formalism result from a specific Ansatz
for HS fields that simplifies the analysis drastically.

The aim of this paper is to explain main elements of the suggested formalism
applying it to the lower-order
vertices. Even at this level the developed formalism allowed us to obtain a new result
 deriving directly
(\ie without by hand field redefinitions as in \cite{Vasiliev:2016xui})
the HS current vertex in the projectively-compact spin-local form of \cite{Vasiliev:2022med}
 with the minimal number of space-time derivatives.

\subsection{Higher-spin equations}
\label{HSsketch1}
The nonlinear equations of \cite{Vasiliev:1992av}
\begin{align}
&\dr_x W+W*W=0\,,\label{HS1}\\
&\dr_x S+W*S+S*W=0\,,\label{HS2}\\
&\dr_x B+[W,B]_*=0\,,\label{HS3}\\
&S*S=i(\theta^{A} \theta_{A}+ B*(\eta \gga +\bar \eta \bar\gga))\,,
\label{HS4}\\
&[S,B]_*=0\,\label{HS5}
\end{align}
are designed
to reproduce  field equations on dynamical HS fields in any gauge and  field variables.
Apart from space-time coordinates $x^n$ with De Rham differential
$\dr_x:= dx^n\f{\p}{\p x^n}$, master fields $W(Z;Y;K|x)$, $S(Z;Y;K|x)$ and $B(Z;Y;K|x)$
 depend on  spinor variables $Z_A=(z_{\al},
\bar z_{\dal})$, $Y_A=(y_{\al}, \bar y_{\dal})$ ($\ga,\gb=1,2$, $\dga,\dgb =1,2$)
and involutive Klein elements
$K=(k,\bar k)$
\be\label{hcom}
\{k,y_{\al}\}=\{k,z_{\al}\}=0\,,\qquad [k,\bar y_{\dal}]=[k,\bar
z_{\dal}]=0\,,\qquad k^2=1\q [k\,,\bar k]=0\,.
\ee
Similarly for $\bar k$.

HS star product is defined as follows
\be\label{star}
(f*g)(Z; Y)=\ff{1}{(2\pi)^4}\int d^4 U d^4 V f(Z+U; Y+U)g(Z-V;
Y+V)\exp(iU_{A}V^{A})\,.
\ee
Indices are raised and lowered with the aid of the symplectic form
$\gep_{AB}=-\gep_{BA}$ as follows: $X^{A}=\gep^{AB}X_{B}$ and
$X_A=X^{B}\gep_{BA}$. (Nonzero elements of $\gep_{AB}$ are $\gep_{\ga\gb}$ and $\gep_{\dga\dgb}$.)
There are two types of anticommuting
differentials in \eqref{HS1}-\eqref{HS5}, namely  space-time
$\mathrm{d}x^{n}$ and  spinor
$\theta^{A}=(\theta^{\al}, \bar{\theta}^{\dal})$. Master fields
have different gradings with respect to these differentials, namely
$W=W_{n}\mathrm{d}x^{n}$ and
$S=S_{\al}\theta^{\al}+\bar{S}_{\dal}\bar\theta^{\dal}$ are one-forms in $dx^n$ and $\theta^A$,
respectively, while $B$ is a zero-form. All products are wedge, with the wedge symbol implicit.
$\theta$-differentials have the following commutation
rules
\be
\{dx^n\,,\theta_{B}\}=0\q
\{\theta_{A}\,,\theta_{B}\}=0\q
\{\theta_{\al},k\}=\{\bar\theta_{\dal},\bar k\}=0\,,\qquad
[\theta_{\al},\bar k]=[\bar\theta_{\dal},k]=0\,.
\ee
Since $\theta^3=\bar\theta^3=0$, that is an equivalent form of the Schouten
identity,
\be\label{klein}
\gga=\exp({iz_{\al}y^{\al}})k\theta^{\al} \theta_{\al}\,,\qquad
\bar\gga=\exp({i\bar{z}_{\dal}\bar{y}^{\dal}})\bar
k\bar\theta^{\dal}\bar\theta_{\dal}\,
\ee
turn out to be central with respect to the star product, \ie
\be
\gga * f = f*\gga\q\bar \gga * f = f*\bar \gga\q\forall f=f(Z;Y;K,\theta)\,.
\ee
%\label{pert}
Following \cite{Vasiliev:1992av}, to analyse equations (\ref{HS1})-(\ref{HS5}) perturbatively  one
starts with the vacuum solution
\be
B_0=0\label{B0}\q
S_0=\theta^\al z_{\al}+\bar{\theta}^{\dal}\bar
z_{\dal}\,.
%\label{S0}
\ee
Plugging it into \eqref{HS1}-\eqref{HS5} and using  that the graded star-product
commutator yields
\be
\label{S0}
[S_0\,,]_* = -2i \mathrm{d}_Z\q \mathrm{d}_Z:= \theta^A \frac{\p}{\p Z^A}\,,
\ee
we find that $W_0$ should
be $Z$-independent, $W_0=\go(Y; K|x)$, and satisfy \eqref{HS1}.
Similarly, at the next order one gets $B_1=C(Y; K|x)$ from $[S_0,
B_1]_*=0$ and $C$ satisfies \eqref{HS3}. This way we find the first
terms on the \rhss of unfolded equations of the form originally proposed in
\cite{Vasiliev:1988sa}
\begin{equation}\label{HSsketch1dr}
\dr_x \omega+\go\ast \go=\Upsilon(\go,\go,C)+\Upsilon(\go,\go,C,C)+\ldots,
\end{equation}
\begin{equation}\label{HSsketch2}
\dr_x C+\omega \ast C-C\ast \omega=\Upsilon(\go,C,C)+\Upsilon(\go,C,C,C)+\ldots\,.
\end{equation}
 As in \cite{Vasiliev:1988sa},
our perturbative
expansion is in powers of the zero-forms $C$ with the one-forms $\go$ treated as
being of order zero.

\subsection{Conventional contracting homotopy analysis}

In the perturbative analysis of the nonlinear HS field equations one starts with
the  $S$-dependent equations (\ref{HS2}), (\ref{HS4}), (\ref{HS5}) on the auxiliary
spinor coordinates $Z^A$ the dependence on which eventually determines the form of the
space-time HS equations  (\ref{HSsketch1dr}), (\ref{HSsketch2})
resulting from  (\ref{HS1}), (\ref{HS3}). As a consequence of (\ref{S0}) the $S$-dependent equations have the form
\be
\label{fg}
\dr_Z f(Z;Y) = g(Z;Y)\,.
\ee
Consistency of the HS equations implies that $g(Z;Y)$ is
$\dr_Z$-closed at every perturbation order,
\be
\dr_Z g(Z;Y)=0\,.
\ee
As a result, equation (\ref{fg}) can be solved in the form
\begin{equation}f={\hmt}_0 g+\mathrm{d}_Z\epsilon+h,\label{eq:gen_sol}
\end{equation}
where $h$ is in $\dr_Z$-cohomology and conventional contracting homotopy
$\hmt_0$ is defined by the Poincar\'e formula
 \begin{equation}
\hmt_0 g\left(Z;Y;\theta\right):=Z^{A}\dfrac{\partial}{\partial\theta^{A}}\intop_{0}^{1}
\dfrac{d t}{t}g\left(t Z;Y;t\theta\right)\,.
\label{eq:dz*}
\end{equation}
 Though looking most natural,  at $\epsilon=h=0$ it does not lead to the local
 (more precisely, spin-local \cite{Gelfond:2019tac}) frame  beyond the free
field level in which case it was used in the
perturbative analysis of HS equations since \cite{Vasiliev:1992av}.
 Generally, the undetermined exact part $\mathrm{d}_Z\epsilon$ and $h$
 describe solutions to the homogeneous equation (\ref{fg}) with $g=0$.
  The freedom in $\epsilon$ is
gauge  while  $h(\go,C)$ with HS fields $\go(Y;K|x)$ and $C(Y;K|x)$
induces nonlinear field redefinitions.

 Shifted homotopy operator  results from the shift
$Z^A\to Z^A+ a^A$ with some $Z$-independent $a^A$.
Shifted contracting homotopy $\hmt_{a}$  and cohomology projector $\hhmt_{a }$ act as follows
\cite{Didenko:2018fgx}
\be\label{homint0} \hmt_{ a} \phi(Z;Y; \theta) =\int_0^1 \f{dt }{t } (Z+ a)^A\f{\p}{\p \theta^A}
 \phi( t  Z-(1-t ) a;Y;t \,\theta)\q\hhmt_{a } \phi(Z;Y; \theta)= \phi(-a;Y;0)\,
\ee
fulfilling the standard resolution of identity
\be\label{newunitres}
\left\{ \dr_Z\,,\hmt_{a }\right\} +\hhmt_{a }=Id\, .
\ee

The form of the resulting equations at $\epsilon=h=0$  depends on the choice of
$\hmt_a$. Transition from one contracting homotopy to another
affects  both $\epsilon$ and $h$. The problem is to identify
a homotopy procedure that leads to the spin-local form of the field equations.

As shown in \cite{Didenko:2018fgx}, in the lowest nonlinear
order locality is achieved with the aid of an appropriate shifted homotopy operator.
This shift is to large extent
determined by the general analysis of \cite{Gelfond:2018vmi}. Further analysis shows
\cite{GV}, that
the extension of these results to the higher orders is not straightforward, suggesting the
 shifts at the higher orders to depend on the homotopy integration parameters like $t $ in
(\ref{homint0})  introduced at lower orders, that looks impossible because
the integration in (\ref{homint0}) has been completed.

The aim of this paper is to work out a novel {\it differential} homotopy procedure
allowing to make homotopy parameter-dependent shifts postponing the integration
to the very last step of the evaluation of  corrections to HS equations. Apart from
 providing a powerful extension of the usual
shifted homotopy procedure,
the proposed scheme leads to a remarkable new representation for the
functions resulting from the application of the differential contracting homotopy to
nonlinear HS equations bringing
the \rhs of equation (\ref{fg}) to a uniform form, that reduces its analysis
to a certain polyhedra cohomology problem with respect to the homotopy  integration
parameters. This  technique provides a far going generalization of the
idea of \cite{Gelfond:2021two} to represent the \rhs of (\ref{fg}) in the manifestly exact form by virtue of
partial integrations and Schouten identity. The remarkable byproduct of our approach is that
the form of the resulting class of functions  accounts both of these ingredients   automatically.

The proposed scheme will be shown to have a number of  remarkable
properties leading to a  significant improvement  of the known results at the
first and second orders in the zero-forms $C(Y)$, that reproduces the
projectively-compact spin-local form of the lower-order vertices  originally obtained by hand
(direct field redefinition) in \cite{Vasiliev:2016xui} and so far  unavailable within the
shifted homotopy approach of \cite{Didenko:2018fgx}-\cite{Didenko:2020bxd}.
Extension of these results to higher orders  will be presented elsewhere.

The rest of the paper is organized as follows. Definition of the differential homotopy is given in Section \ref{difhom} where also  the fundamental Ansatz
that results from the application of the differential homotopy approach and
captures various shift parameters of \cite{Didenko:2018fgx,Didenko:2019xzz} is presented
and the  duality between the HS homological problem
and that in the  homotopy parameter space is  established. The star-multiplication
formulae analogous to the star exchange formulae of \cite{Didenko:2018fgx} are presented in
Section \ref{Star multiplication f}. Special properties of the fundamental
Ansatz for particular values of the  homotopy parameters  are discussed in Section \ref{spec}.
The differential homotopy form of the  nonlinear system of HS equations is
presented in Section \ref{hdeq}. Examples of  application of the proposed
approach to the derivation of lower-order expressions for HS fields and equations
are collected in Section \ref{LOC}. General lessons of the lower-order analysis are
summarized in Section \ref{less} including the Ansatz for the
projectively-compact vertices.
Conclusions and perspectives are  Section \ref{con}.

\section{Differential homotopy}
\label{difhom}
\subsection{Total differential}

The homotopy   parameter $t$ in (\ref{homint0}) and its further extension $t^a$
 are  interpreted as additional coordinates of some manifold $M$ with the total differential
\be
\label{td}
\dr:= \dr_Z +\dr_t \q
\dr_Z := \theta^A \f{\p}{\p Z^A}\q \dr_t : = dt^a\f{\p}{\p t ^a}\,,
\ee
where $\theta^A$ and $dt^a$ are anticommuting differentials.
It is assumed that $M =\M_Z\times \M$ with some compact homotopy space
$\M$ and  $\M_Z=\mathbb{R}^4$. Integration in $M$ will be over
$\M$ or, more generally (though equivalently),  a $\dim \M$-dimensional compact cycle
in $M$.
%(Later on we shall see that this construction is useful to extend to the
%noncommutative spaces associated with the integration variables like $U^A$ and $V^A$ in
%(\ref{star}).)

Equations to be solved at every perturbation order are of the form
\be
\label{totg}
\dr f(Z,t ,\theta,dt ) = g(Z,t ,\theta,dt )\,,
\ee
with some $\dr$-closed $g$,
\be
\dr g(Z,t ,\theta,dt )=0\,.
\ee
Homotopy coordinates $t ^a$, $a=1,\ldots n$ vary in a compact domain
of $\mathbb{R}^n$:
functions like $f$ and $g$ contain $\theta$- and $\delta$- functions like
$\theta(t )\theta(1-t )$ confining  integration to a compact $\M$.

Physical fields and equations in HS theory
are  supported by the $\dr $ cohomology  which, in the case of
 compact $\M$ relevant to our analysis, is carried by the $Z^A$,
$\theta^A$--independent integrals over $\M$ with $\dr_t $ cohomology
represented by the volume integral.  ($H^0(\dr_t)$   plays no role in our analysis
while $H^p(\dr_t)=0$ at $0< p < \dim \M$.)

The idea of differential homotopy is based on the removal of the integrals in the
equations
\be
\dr_Z f_{int} = g_{int}\q g_{int} = \int_{\M } g(Z,\theta, t ,dt)\q
f_{int} = \int_{\M} f(Z,\theta, t ,dt)\,,
\ee
resulting in
\be
\dr_Z f = g +\dr_t  h +g^{weak}\,,
\ee
where $g^{weak}$ does not contribute to the integral because its degree $\deg g^{weak}$ as
a form in $\M$ differs
from $\dim \M$. Setting $g^{weak} = \dr_Z h - \dr_t f $ (taking into account that
$\deg h=\dim \M-1$ and $\deg f =\dim \M$) and replacing $f\to f-h$, we obtain (\ref{totg}).

Equations (\ref{totg}) are invariant under diffeomorphisms in the total space with
local coordinates $Z^A, t ^a$ and differentials $\theta^A, dt ^a$\,.
Moreover, sometimes it is convenient to extend the setup by the
coordinates like $U^A$ and $V^A$ of the  non-compact non-commutative space (\ref{star}),
\be
\label{tdUV}
\dr:= \dr_Z +\dr_t +\dr_U+\dr_V+\ldots \q
\dr_U:= dU^A\f{\p}{\p U^A} \q \dr_V:= dV^A\f{\p}{\p V^A}\,.
\ee

\subsection{Integration}
\label{int}

To avoid a sign ambiguity due to (anti)commutativity of
differential forms we recall that every differential expression
is in the end accompanied with a set of integrals on the left
\be
\int_{t ^1}\ldots \int_{t^k} f(t ,dt )\,.
\ee
We take the convention that
\be
\label{sign1}
\int_{t^n} \int_{t^m}\mu(t, dt) = - \int_{t^m} \int_{t^n}\mu(t, dt)
\ee
and
$$
 \int_t  dt  \delta(t -a
  ) = 1\,.
$$
The latter formula will also be used  other way around inserting a unity in the form of integral
of a delta-function of a new variable.

To simplify formulae  we will  use a shorthand notation
\be
\label{short}
 \int_{t^1}\ldots \int_{t^k}\to \int_{t^1\ldots t^k}
\ee
with the convention that $\int_{t^1\ldots t^k}$ is totally antisymmetric
in $t^a$. (This  mimics the Levi-Civita symbol in the
component form of the integral of a differential form.)
Also, we will use the  convention that, whenever convenient, $\int_{t^1\ldots t^k}$ can
be written anywhere in the expression for the differential form to be integrated
at the condition
\be\label{sign2}
\dr \int_{t^1\ldots t^k} = (-1)^k \int_{t^1\ldots t^k} \dr \q
\lambda \int_{t^1\ldots t^k} = (-1)^{kp} \int_{t^1\ldots t^k} \lambda
\ee
for any $p$-form $\lambda$. Though we do not perform integrations till the
very last step of the vertex computation, it is useful to keep the integral
 symbol
to control sign factors resulting from (\ref{sign1}) and (\ref{sign2}).

\subsection{Fundamental Ansatz}
\label{Ansatz}
%\subsection{Ansatz}

Being originally designed to capture a most general class of solutions of the HS equations, the
differential homotopy approach turns out to be useful in many more respects as we explain now
starting from the simplest case of the lowest order holomorphic deformation linear in $\eta$.

As shown in Section \ref{LOC}, direct  lower-order computation in the holomorphic sector
within the differential homotopy approach
 yields  expressions of the following remarkable form
\be
\label{fbeta}
f_\mu=  \!\int_{p^2_i r^2_i u^2 v^2 \tau\gs\gb\rho }\ls\mu(\tau,\gs,\gb,\rho,u,v, p,r )
\dr \Go^2
\Ex G (g(r)) \,,
\ee
where
\be
\dr \Go^2 := \dr \Go^\ga \dr \Go_\ga\,,
\ee
\be
\label{sW}
 \Go_\ga (\tau,\gs,\gb,\rho,u,v,p ):= \tau z_\ga - (1-\tau) (p_\ga (\gs) -\gb v_\ga +\rho(  y_\ga +p_{+\ga} +u_\ga )) \,,
\ee
\be
\label{p+=}
p_{+\ga} := \sum_{i=1}^l p_{i\ga}
%\q p_{j\ga} = -i \frac{\p}{\p r_j^\ga}\,
\ee
and
\be
p_\ga(\gs):=\sum_{i=1}^l p_{i\ga} \gs_i
\ee
with some integration parameters $\gs_i$,
\be\label{Exp}
\Ex:=   \exp i \big ( \Go_\gb(y^\gb +p^\gb_++u^\gb)+u_\ga v^\ga - \sum_{i=1}^l p_{i\ga} r_i^\ga
-\!\sum_{l\geq j>i\geq 1}p_{i\gb}
p_j^\gb \big)\,,
\ee
\be
\label{Gk}
G_l(g):= g_1(r_1)\ldots g_l(r_l)k \,,
\ee
$g_i(y)$ are some functions of $y_\ga$ (e.g., $C(y)$ or $\go(y)$). Antiholomorphic variables $\bar y_\dga$, Klein operators
$K$, space-time coordinates $x$ in the arguments of $C(Y;K|x)$ and
the antiholomorphic star product $\bar *$ between the product factors $g_i$ are implicit).

 Analogously to (\ref{tdUV}) in the sequel we will assume that
\be
d^2 u :=du^\ga du_{\ga}\q d^2 v :=dv^\ga dv_{\ga}\q
d^2 p_i :=dp_i^\ga dp_{i\ga}\q d^2 r_i :=dr_i^\ga dr_{i\ga}\,,
\ee
where $du^\ga$, $dv^\ga$, $dp_i^\ga$ and  $dr_i^\ga$ are anticommuting differentials treated
on equal footing with $\theta^A, d\tau$, $d\gs_i$, $d\rho$ and  $d\gb$. In particular,
\be
\dr =\dr_Z+d\tau\f{\p}{\p \tau}
 +d\rho\f{\p}{\p \rho}+ d\gs_i \f{\p}{\p \gs_i} +
d\gb\f{\p}{\p \gb} +du^\ga\f{\p}{\p u^\ga} + dv^\ga\f{\p}{\p v^\ga}\,
+dp_i^\ga\f{\p}{\p p_i^\ga} + dr_i^\ga\f{\p}{\p r_i^\ga}
\ee
and
\be
\label{muv}
\mu(\tau,\gs,\gb,\rho,u,v,p,r ) =\mu(\tau,\gs,\gb,\rho)d^2 u d^2 v \prod_{i=1}^l
d^2 p_i d^2 r_i\,.
\ee
We use  such a normalization  that
\be
\label{meas}
\qquad\int d^2 u d^2 v \exp i u_\gb v^\gb =1\,.
\ee

 Correspondingly,
\be
\label{dWGo}
\dr\Go_\ga (\tau,\gs,\gb,\rho ) = \tau \theta_\ga +d\tau (z_\ga + p_\ga (\gs) -\gb v_\ga +\rho(  y_\ga +p_{+\ga} +u_\ga ))
-(1-\tau) \big(\sum_i p_{i\ga} d\gs_i -d \gb v_\ga +\dr \rho (  y_\ga +p_{+\ga} +u_\ga )\big )
\,,
\ee
where the dependence on $u_\ga,v_\ga$ and $p^\ga_i$ on the \lhs is implicit and
the terms with  $du_\ga$, $dp_i^\ga$ and $dv_\ga$ have been discarded since they do not
contribute upon multiplication by the measure factor $ d^2 u d^2 v \prod_{i=1}^l
d^2 p_i d^2 r_i$ in (\ref{muv}). (Note that though such terms can be treated as
weak they are a kind of trivial never contributing to the final result as long as the
measure factor $ d^2 u d^2 v \prod_{i=1}^l d^2 p_i d^2 r_i$ is preserved at all stages.)

It is convenient to assume that the integration over homotopy parameters
like $\tau$, $\gs_i$, $\rho$, $\gb$ is over $\mathbb{R}^m$ of an appropriate dimension
$m$ while $\mu(\tau,\gs,\gb,\rho)$ has a compact support $\M$ determined by  $\delta$ and
$\theta$ functions of some  combinations of the homotopy parameters.
We shall see that in practice $\mu(\tau,\gs,\gb,\rho)$ is supported by certain polyhedra.

Note that,  by virtue of  Taylor expansion
$g_i(v_i) = \exp [ v_i^\ga\f{\p}{\p y^{\ga}}] g_i(y)\big |_{y=0}$, formulae (\ref{fbeta}), (\ref{Exp})  can
be rewritten in the differential form with $r_i^\ga$ integrated out to give
\be
\label{pd}
p_{j\ga} =-i \f{\p}{\p r^\ga_j}
\ee
at $r_j^\ga=0$ upon differentiation. In the sequel, this differential form will be used as well.

It should be stressed that $\M$ is demanded to be compact to guarantee that the generating functions like (\ref{fbeta})
are entire in $p_{j\ga} $, \ie derivatives with respect to spinor
variables which, by virtue of the unfolded HS equations discussed in Section \ref{LOC},
in turn implies that the resulting space-time
HS equations are expandable in space-time derivatives.

The differentials $\dr\Go_\ga$ and therefore $\dr\Go^\ga \dr\Go_\ga$ contain various combinations
of $\theta^\ga$ and $dt^a=(d\tau, d\gs_i, d\rho,d\gb)$. Hence not all of the
terms in  $\dr\Go^2$ contribute
to the  integrated expression. In particular, $\dr\Go^\ga \dr\Go_\ga$ always contains the
$\theta^\ga \theta_\ga$ term. To put it differently, the measure $\mu \dr\Go^2$ may contain
weak terms that do not contribute upon  integration. A related comment is that $\mu(\tau,\gs,\gb,\rho)$ in  (\ref{muv}) may not contain the differentials for all its
arguments.
We will come back to this issue in  Section \ref{hdeq}.

The  parameters $\gs_i$, $\gb$ and $\rho$ are related to the shift parameters of
\cite{Didenko:2018fgx, Didenko:2019xzz}. The
difference is that in our approach these are integration variables while in
\cite{Didenko:2018fgx, Didenko:2019xzz} parameters like
$\gs_i$ and $\gb$ were taking definite values, say, $\gs_{0i}$ and $\gb_{0}$. In fact, this difference is
not too important since in the new approach one can choose $\mu(t_a)$
to contain the factors of
$d\rho \delta(\rho-\rho_0)$ and/or $d\gb \delta(\gb-\gb_0)$. Other way around, the results of
\cite{Didenko:2018fgx, Didenko:2019xzz}
at a given perturbation order can be integrated over the homotopy parameters
like $\gb_0$ and $\rho_0$, that appear at the same order.

The most important feature of the Ansatz (\ref{fbeta}) is that the homotopy parameters
$\tau$, $\gs_i$, $\gb$ and $\rho$ contribute either via the measure
$\mu(\tau,\gs,\gb,\rho,u,v, p,r )$ or via $\Omega$ (\ref{sW}). As we explain now, this
reduces the cohomological HS problem to that on the homotopy space $\M$.

\subsection{Homology map}
\label{cohmap}
Formula (\ref{fbeta}) has the remarkable property that its $\Go$-dependent part is $\dr$-closed,
\be
%d^2 u d^2 v \prod_{i=1}^l
%d^2 p_i d^2 r_i
%\dr[(\dr\Go)^2\Ex]=
d^2 u d^2 v \prod_{i=1}^l
d^2 p_i d^2 r_i\,\dr \Big ( \dr \Go^2
\exp i \big ( \Go_\gb(y^\gb +p^\gb_++u^\gb)+u_\ga v^\ga
-\! \!\sum_{i=1}^l p_{i\ga} r_i^\ga-\ls\sum_{k\geq j>i\geq 1}
\! p_{i\gb}
p_j^\gb \big)\Big ) =0\,.
\ee
This is because the one-forms $\dr \Go^\ga$ are anticommuting and hence,
analogously to
$\theta^3=0$,
\be
\label{sch}
(\dr\Go)^3=0
\ee
 since $\ga=1,2$. (Analogously the terms with $dr_i$ and $dp_i$ do not contribute because
 $(dr_i)^3=(dp_i)^3=0$.)

As a result, the differential $\dr$, that acts on $f_\mu$ (\ref{fbeta}), effectively acts
only on the measure $\mu$ on the \rhs of (\ref{muv}),
\be
\label{dmu}
\dr f_\mu = (-1)^N f_{\dr \mu}\,,
\ee
where $N$ is the number of integration parameters $\tau, \gs_i, \gb,\rho$.
This maps the original homological problem in terms of $Z$ variables to that
on the space of the measure factors $\mu(t_i)$ with no reference
to the spinor variables. The great advantage of this formalism is that there is no
 need
to use the Schouten identity in practical computations: the only formula where
Schouten identity in the form (\ref{sch}) manifests itself is (\ref{dmu}).
(The necessity of accounting the Schouten identity in the formalisms available so far was
one of the hardest complications.)

By virtue of (\ref{dmu}),  equation (\ref{totg}) amounts to
\be
f_{\dr \mu_f} = g_{\mu_g}\,.
\ee
This demands
\be
\dr \mu_f \cong \mu_g\,,
\ee
where  $\cong$ denotes the weak equality up to possible weak terms, that do not contribute
under the integrals in $f_{\dr \mu_f}$ and $ g_{\mu_g}$ because the number of differentials
does not match the number of the integration variables. That $g$ in (\ref{totg}) is
$\dr$ closed implies that $\mu_g$ must be weakly $\dr$-closed,
\be
\dr \mu_g\cong 0\,.
\ee
In most cases this implies that
\be
\mu_g \cong \dr h_g
\ee
allowing to set
\be
\mu_f = h_g\,.
\ee

Being analogous to the approach of \cite{Gelfond:2021two}, where the equations $\dr_Z f = g$ were
resolved explicitly in terms of the preexponent factors that however  resulted in a rather involved
procedure due to the lack of the formula (\ref{dmu}) and necessity to resolve the mess resulting from
the straightforward application of  the Schouten identity, the new approach is incomparably simpler as will
be illustrated in the examples of Section \ref{LOC}. The same time it is far more general allowing a broader class of differential homotopy procedures.

\subsection{General homotopy}
\label{General}

Let $f_\mu$ be of the form (\ref{fbeta}), (\ref{sW}). Consider $g'$ of the form
\be\label{g'}
g'_{\mu'}= \int_{p^2_i r^2_i u^2 v^2,\tau'\tau\gs\gb\rho } \ls\ls\ls\ls\ls l(\tau')
  \mu'(\tau',\tau,\gs,\gb,\rho, p,r,u,v ) \dr \Go'{}^2
\Exp
G_k(g(r))\,,
\ee
where $l(\tau)$ is the characteristic function of the unit segment $l(\tau)=1$ for
 $\tau\in [0,1]$ and zero otherwise,
\be
\label{lt}
l(\tau):= \theta(\tau)\theta(1-\tau),
\ee
with some weakly closed $\mu'(\tau',\tau,\gs,\gb,\rho, p,r,u,v )$,
\be\label{dmu'}
\dr\mu'(\tau',\tau,\gs,\gb,\rho, p,r,u,v )\cong 0\,,
\ee
such that
\be\label{mu1}
\mu'(\tau',\tau,\gs,\gb,\rho, p,r,u,v )\Big |_{\tau'=1,d\tau' =0}= \mu(\tau,\gs,\gb,\rho, p,r,u,v )
 \,,
\ee
\bee
\label{sW'}
\Go'_\ga (\tau',\tau,\gs,\gb,\rho ) :=&& \ls\tau'\tau z_\ga - (1-\tau)
\big ( p_\ga (\gs) -\gb v_\ga +
\rho(y_\ga +p_{+\ga} +u_\ga )\big) \nn\\&&\qquad \!\!
-(1-\tau') \big (p_\ga (\gs') -\gb' v_\ga +\rho'
(  y_\ga +p_{+\ga} +u_\ga ) \big)\,,
\eee
where
\be
\gs_i'= \gs_i'(\tau,\tau', \gs,\gb,\rho)\q
\gb'= \gb'(\tau,\tau', \gs,\gb,\rho)\q
\rho'= \rho'(\tau,\tau', \gs,\gb,\rho)\,
\ee
and
the additional term with $(1-\tau')$ is demanded to vanish at $\tau'=1$,
\ie $\gs'_i$, $\gb'$ and $\rho'$ should not be too singular in $1-\tau'$.
Naively, formulae (\ref{dmu'}) and (\ref{mu1}) hold true for
$\mu'(\tau',\tau,\gs,\gb,\rho ) = \mu(\tau,\gs,\gb,\rho ) $ since $\mu$ has to be $\dr$-closed.
One has to be careful however at this point
since $\dr \mu$ is weakly zero $\dr \mu \cong0$ and,
as will be illustrated in some more detail in Section \ref{LOC}, this property can be lost in presence
of the new integration parameter $\tau'$.

Since the differentiation of $g'$ (\ref{g'}) only hits $l(\tau')$,
we obtain
\be
\label{hom'}
\dr g'_{\mu'}  = f_\mu -h_{\mu'}\,,
\ee
where
\be
h_{\mu'} =\int_{\tau'} \De(\tau')\int_{p_i^2 r_i^2 u^2 v^2\tau\gs\gb\rho } \ls
  \mu'(\tau',\tau,\gs,\gb,\rho,p,r,u,v ) (\dr \Go')^2 \Exp G(g) \big |_{r=0}\,,
\ee
with the notation often used in the sequel,
\be
\label{De}
\De (a) :=d a \delta(a)\,.
\ee
Formula (\ref{hom'}) represents $f_\mu$ as a combination
of the $\dr$-exact and cohomological term just as in the standard Poincar\'e homotopy formula.

Finally, let us note that $\Go'$ (\ref{sW'}) can be represented in the form (\ref{sW}) with
\be
\label{sW'eff}
\Go'_\ga (\tau',\tau,\gs,\gb,\rho ) := \tilde \tau z_\ga - (1-\tilde \tau) ( p_\ga (\tilde \gs) -
\tilde \gb v_\ga +\tilde
\rho(  y_\ga +p_{+\ga} +u_\ga ) )\,,
\ee
where
\be
\tilde \tau := \tau' \tau\,,
\ee
\be
\tilde \gs_i  := \frac{1-\tau}{1-\tilde \tau }\gs_i + \frac{1-\tau'}{1-\tilde \tau } \gs'_i\q
\tilde \gb := \frac{1-\tau}{1-\tilde \tau }\gb  +\frac{1- \tau'}{1-\tilde \tau } \gb'\q
\tilde \rho := \frac{1-\tau}{1-\tilde \tau } \rho +\frac{1- \tau'}{1-\tilde \tau } \rho'\,.
\ee
It is important that all these expressions are regular since, for $\tau,\tau'\in [0,1]$,
\be
\frac{1-\tau}{1-\tilde \tau }\in [0,1]\q \frac{1-\tau'}{1-\tilde \tau }\in [0,1]\,.
\ee

\section{Star multiplication formulae}
\label{Star multiplication f}
The so-called star-exchange formulae of \cite{Didenko:2018fgx} for
the star multiplication of the result of the action of the contracting homotopy operator with a
$Z$-independent function greatly simplify the perturbative analysis of HS theory.
The  differential homotopy approach proposed in this paper
admits their useful analogue presented in this section.
Note that the Ansatz with parameters
$\gs_i$, $\rho$ and $\gb$ captures all types of shifted homotopy of
\cite{Didenko:2018fgx, Didenko:2019xzz}.

For convenience, we confine ourselves to the holomorphic sector with $Z$ and $Y$ variables
carrying only undotted indices. The antiholomorphic sector of dotted indices can be treated analogously.

For $f_\mu$ (\ref{fbeta}) the star products with a $Z$-independent function $\varphi(y)$ yield
\be
\label{left}
\varphi(y)*f_\mu=
\int_{p^2_i r^2_i u^2 v^2\tau\gs\gb \rho\gs_\varphi }\ls\De(\gs_\varphi+(1-\gb))
\mu(\tau,\gs,\gb,\rho,p,r,u,v ) \dr \Go^2
\Ex
G_{k+1} (\varphi,g) \,,
\ee
where $\Go_\ga$ still has the form (\ref{sW}) with
$p_{+\ga}$ now including
$
p_\varphi \equiv p_{0\ga},
$
and
\be
\label{right}
f_\mu*\varphi(y)=
\int_{p^2_i r^2_i u^2 v^2\tau\gs\gb\rho\gs_\varphi }\ls\De(\gs_\varphi-(1-\gb))
\mu(\tau,\gs,\gb,\rho,p,r,u,v ) \dr \Go^2
\Ex G_{k+1}(g,\varphi)\,
\ee
with $\Go_\ga$ of the form (\ref{sW}) with $p_{k+1\,,\ga}\equiv  p_{\varphi\,\ga}$.

Let us explain for instance how formula (\ref{right}) is derived. Star product
(\ref{star}) yields
\be
\label{right1}
f_\mu*\varphi(y)=
\int_{p^2_i r^2_i s^2 t^2 u^2 v^2\tau\gs\gb\rho }\ls\ls\ls\ls \ls
d^2s d^2 t \mu(\tau,\gs,\gb,\rho, p,r,u,v ) \dr \Go^2 \exp i[s_\ga t^\ga ]
\Ex g_1(r_1)\ldots g_l(r_l) \varphi (-y-t) k
 \,,
\ee
where
\be\label{Exp1}
\Ex:=   \exp i \big ( \Go_\gb(y^\gb+ s^\gb +p^\gb_++u^\gb)+u_\ga v^\ga - \sum_{i=1}^l p_{i\ga} r_i^\ga
-\!\sum_{k\geq j>i\geq 1}p_{i\gb}
p_j^\gb \big)\,,
\ee
\be
\Omega_\ga = \tau (z_\ga +s_\ga) - (1-\tau) \big (p_\ga (\gs) -\gb v_\ga +\rho (y_\ga +
s_\ga + p_{+\ga} +u_\ga) \big )\,.
\ee

Note that the sign  of the argument of $\varphi (-y-t)$ is changed because the Klein operator $k$
has been moved to the right. Rewriting $\Omega_\ga$ in the form
\be
\Omega_\ga = s_\ga + \Omega'_\ga
\ee
with
\be
\Omega'_\ga = \tau z_\ga  - (1-\tau) \big (p_\ga (\gs) +s_\ga -\gb v_\ga +\rho (y_\ga +
s_\ga + p_{+\ga} +u_\ga) \big )\,
\ee
and using that
\be
\dr \Omega_\ga = \dr \Omega'_\ga
\ee
since $\dr s_\ga$ does not contribute to (\ref{right1}), we obtain after renaming
$t_\ga = -y_\ga - r_{l+1\ga}$
 \bee
\label{right12}
f_\mu*\varphi(y)&&\ls=
\int_{p^2_i r^2_i s^2 u^2 v^2\tau\gs\gb\rho }\ls\ls\ls\ls \ls
d^2s d^2 t \mu(\tau,\gs,\gb,\rho, p,r,u,v ) \dr \Go'^2 \exp i\Big[
u_\ga v^\ga-s_\ga (r_{l+1}^\ga +y^\ga)
+s_\ga (y^\ga + p_+^\ga +u^\ga) \nn\\&&  \!-\sum_{i=1}^l p_{i\ga} r^\ga_i
+\Omega'_\gb (y^\gb + s^\gb +p_+^\gb +u^\gb) -\sum_{ l \geq j>i\geq 1} p_{i\ga } p_j^\ga
\Big ]
 g_1(r_1)\ldots g_l(r_l) \varphi (r_{l+1}) k
 \,.
\eee
Finally, setting $s_\ga = p_{l+1 \ga}$, shifting $v_\ga \to v_\ga +p_{l+1\ga}$, and renaming
$\Omega'_\ga \to \Omega_\ga$ we recover (\ref{right}). Formula (\ref{left}) is derived
analogously.

Consider now the case where the star product with a $Z$-independent $\varphi(y)$
acts on one of the inner factors of $g_l(y)$. For instance, let
\be
g_l(y)\to g_l(y) *\varphi (y)\,.
\ee
Then it is easy to see that, due to translational invariance of the star product,
$p_l$ gets replaced by $p_l+p_\varphi$ and star product adds a
factor of $\exp[-i p_{l\ga} p_\varphi^\ga]$. As a result,
\be
\label{lr}
\ls f_\mu (g_l*\varphi)=
\int_{\tau\gs\gb\rho p_i^2 r_i^2 u^2 v^2 \gs_\varphi }\,\ls\ls\ls\ls\ls\De(\gs_\varphi-\gs_l)
\mu(\tau,\gs,\gb,\rho, p,r,u,v) \dr \Go^2
\Ex G(\ldots g_l(r_l), g_\varphi (r_\varphi ), g_{l+1} (r_{l+1})\ldots )
\,.
\ee

Analogously, for
\be
g_{l+1}(y)\to \varphi (y)* g_{l+1}(y) \,,
\ee
\be
\label{l1l}
\ls f_\mu (\varphi*g_{l+1})=
\int_{\tau\gs\gb\rho p_i^2 r_i^2, u^2 v^2 \gs_\varphi }\ls\ls\ls\ls\ls\De(\gs_\varphi-\gs_{l+1})
\mu(\tau,\gs,\gb,\rho,p,r,u,v ) \dr\Go^2
\Ex
G( \ldots g_l(r_l), g_\varphi (r_\varphi ), g_{l+1} (r_{l+1})\ldots)\,.
\ee
(It is  convenient to assume  that the label $\varphi$ of $p_{\varphi\ga}$ is between
$l$ and $l+1$.)

At the condition that, in agreement with the results of \cite{Didenko:2019xzz},
$-\infty <\gb< 1$, all variables $\gs_i$ vary from $\gb-1$ to $1-\gb$,
\be
\gs_i\in [\gb-1,1-\gb]\,,
\ee
formulae (\ref{left}), (\ref{right}), (\ref{lr}) and (\ref{l1l}) have the following important consequences
used in Section \ref{LOC} that, in fact, relate the Hochschild cohomology of the HS problem to the De Rham cohomology of
the measure factors of $\mu$ on the compact polyhedra,
\be
\label{leftc}
\varphi(y)*f_\mu-f_\mu(\varphi*g_1) =
\int_{\tau\gs\gb \rho\gs_\varphi }\ls\ls\dr P (\overleftarrow{\gs},\gs_\varphi,\gs_1)
\mu(\tau,\gs,\gb,\rho ) \dr \Go^2
\Ex
G(\varphi,  g_1\ldots )\,,
\ee
\be
\label{rightc}
f_\mu(g_k*\varphi)-f_\mu*\varphi(y) =
\int_{\tau\gs\gb \rho\gs_\varphi }\ls\dr P (\gs_k,\gs_\varphi, \overrightarrow{\gs})
\mu(\tau,\gs,\gb,\rho ) \dr \Go^2
\Ex
G( \ldots g_k , \varphi ) \,,
\ee
\be
\label{midl}
f_\mu(\varphi*g_{l+1})-f_\mu(g_l*\varphi) =
\int_{\tau\gs\gb \rho\gs_\varphi }\ls\ls\dr P(\gs_l,\gs_\varphi,\gs_{l+1})
\mu(\tau,\gs,\gb,\rho ) \dr \Go^2
\Ex G(\ldots g_l , \varphi , g_{l+1}\ldots )\,,
\ee
where the spinor integration variables $p_i^\ga$, $r_i^\ga$, $u^\ga$ and $v^\ga$ are implicit,
\be
\label{boundb}
\overleftarrow{\gs} = \gb-1 \q \overrightarrow{\gs} = 1-\gb\,
\ee
and
\be
P(\gs_l,\gs_{l+1},\gs_{l+2}):=\theta(\gs_{l+2}-\gs_{l+1})\theta(\gs_{l+1}-\gs_l)\,.
\ee

Later on we will see that the extension of $P(\gs_l,\gs_{l+1},\gs_{l+2})$ to an
arbitrary string of arguments
\be
\label{Kk}
P_k(\gs_l,\gs_{l+1},\ldots \,,\gs_{l+k}):=\theta(\gs_{l+k}-\gs_{l+k-1})\ldots
\theta(\gs_{l+1}-\gs_l)
\ee
also plays a role in the analysis. In these terms, $P(\gs_l,\gs_{l+1},\gs_{l+2})$ is
$P_2(\gs_l,\gs_{l+1},\gs_{l+2})$.

To summarize, the remarkable property of formulae (\ref{fbeta}), (\ref{sW}) is that they
preserve  their form
under the star product with  $Z$-independent functions.

\section{Special properties}
\label{spec}

Written in terms of differential forms, formula (\ref{fbeta}) is invariant under diffeomorphisms
in the homotopy  variables $t^a\to t'{}^a  (t)$. In most cases such diffeomorphisms affect
the form of $\Go_\ga$ (\ref{sW}) and, hence,  the exponent (\ref{Exp}) in (\ref{fbeta}). These
can be applied of course, but it does not help  just affecting the form of the expression
as a whole.
However, remarkably, the dependence on
all coordinates $\gs_i,\gb,\rho$ disappears from $\Go^\ga$ at $\tau=1$ and the dependence
on $\rho $
disappears from the exponent  in (\ref{fbeta}) since
$$(y_\ga+u_\ga+ p_{+\ga}) (y^\ga+u^\ga +p^\ga_+)=0\,.$$
This has two consequences allowing to perform any change of variables in $t^a$ at $\tau=1$
and in $\rho$ at any $\tau$    without
affecting the exponent (\ref{Exp}). Let us consider these important cases in more detail.

\subsection{$\tau=1$}

Let $t^a$ denote all homotopy coordinates except for $\tau$. (For simplicity we
discard the sector of antiholomorphic variables.)
Let the measure factor $\mu(\tau,t)$ be of the form
\be
\mu(\tau,t, d\tau ,dt ) = \De (1-\tau) \tilde \mu(t,dt)\,.
\ee
In that
case, the $t^a$-dependent terms drop out from the exponent in (\ref{fbeta}) while the
$t^a$, $d t^a$-dependent terms drop out from $\dr\Go^\ga \dr\Go_\ga$
 with $\Go_\ga$ (\ref{sW}),
 \be
\De(1-\tau )\Go^\ga = \De(1-\tau ) z^\ga\q \De(1-\tau )\dr \Go^\ga = \De(1-\tau ) \theta^\ga\,.
\ee
  This allows us to make any change of variables $t^a$ which only affect the
measure  $\mu$. Such changes of variables can be done
freely at any stage of the computation.

\subsection{$\rho$}

That the dependence on $\rho $ drops out  from the exponent in (\ref{fbeta})
is the manifestation and further extension of the phenomenon
found in \cite{Didenko:2018fgx,Tarusov:2022qpo} at $\gb=0$ of the independence
of  vertices bilinear in $C$ on the uniform shifts
\be
\gs_i\to\gs_i + \lambda\q \rho\to \rho+\lambda
\ee
in the variables of those references which is equivalent to
\be
\gs_i\to\gs_i \q \rho\to \rho+\lambda
\ee
in the variables of this paper. In \cite{Didenko:2018fgx,Tarusov:2022qpo}
 the parameter $\rho$ was a constant while in this paper it is a variable, \ie a
 coordinate in $\M$.
Though the dependence on $\rho $ drops out from the exponent in (\ref{fbeta}),
it contributes to $\dr\Go^2 $ in a very special way.

 Indeed, let us set
\be
\label{sw+}
\Go_\ga (\tau,\gs,\gb,\rho) = w_\ga (\tau,\gs,\gb) - (1-\tau) \rho (y+u+p_+)_\ga
  \,,
\ee
\be
\label{w}
w_\ga (\tau,\gs,\gb ) := \tau z_\ga - (1-\tau) (p_\ga (\gs) -\gb v_\ga )\,.
\ee
Using the differential realization of $p_{j\ga}$ (\ref{pd}), this yields
\be
\label{fbetaw}
f_\mu=  \!\int_{ u^2 v^2 \tau\gs\gb\rho }\ls\mu(\tau,\gs,\gb,\rho,u,v )
\dr \Go^2
\Ex g_1(r_1)\ldots g_k(r_k) \big |_{r_l=0}k \,
\ee
with
\be\label{Expw}
\Ex:=   \exp i \big ( w_\gb(y^\gb +p^\gb_++u^\gb)+u_\ga v^\ga
-\!\sum_{k\geq j>i\geq 1}p_{i\gb}
p_j^\gb \big)\,.
\ee
{}From (\ref{sw+}) it follows that
\be\label{dOm}
\dr \Go^2 = \dr w^2 -2 \dr ((1-\tau)\rho) (y^\ga+u^\ga+p_+^\ga) \dr  w_\ga\,
\ee
and, hence,
\be
\label{fbetawd}
f_\mu=  \!\int_{ u^2 v^2 \tau\gs\gb\rho }\ls\mu(\tau,\gs,\gb,\rho,u,v )
\big (\dr w^\ga \dr w_\ga +2i \dr ((1-\tau)\rho) \dr\big )
\Ex g_1(r_1)\ldots g_k(r_k) \big |_{r_l=0}k \,.
\ee

Since any change of variables of the form
\be
\label{frr}
\rho \to \rho '(\rho, \tilde t^a)\q \tilde t'{}^a= \tilde t^a\q \tilde t^a =
(\tau, \gs_i,\gb) \,
\ee
does not affect the exponent, it can be freely performed only
affecting  the form of the factor $\mu \dr \Go^\ga \dr \Go_\ga$
to relate seemingly different expressions.
This property will play instrumental role in the analysis of Section \ref{B2}
of the form of field equations guaranteeing equivalence of the spinor
and space-time spin-locality according to the
criterion of \cite{Vasiliev:2022med}.

\subsection{$\gb$}
\label{gb}

An interesting feature of the $\gb$-dependence of the shifted homotopy procedure
is that it does not affect the lower-order analysis as was
explained in \cite{Didenko:2019xzz} in terms of  certain star-product re-ordering  inspired by the analysis of \cite{DeFilippi:2019jqq,DeFilippi:2021xon}. Here we  explain it in a somewhat different way
within the approach of this paper.

Consider the $\gb$-dependent part of the Ansatz (\ref{fbeta}),
\be
\label{anf}
X=\int_{u^2 v^2\tau\gs\gb\rho }\ls\mu(\tau,\gs,\gb,\rho,u,v ) \dr \Go^2
\exp i [u_\ga v^\ga+\Go_\ga (y^\ga +p^\ga_+ + u^\ga) ]\,,
\ee
with $\Go_\ga$ (\ref{sW}) and
$
\mu(\tau,\gs,\gb,\rho,u,v) =\mu(\tau,\gs,\gb,\rho)d^2 u d^2 v \,.
$
 This yields
\bee
X=&& \ls\int_{u^2 v^2 \tau\gs\gb\rho }\ls\mu(\tau,\gs,\gb,\rho,u,v ) \dr \Go^2
\exp i[u_\ga v^\ga + ( (1-\tau) \gb v_\ga +\Go^0_\ga )(u^\ga +y^\ga +p_+^\ga)]\\&&
\ls \ls= \int_{u^2 v^2 \tau\gs\gb\rho }\ls\mu(\tau,\gs,\gb,\rho, u,v ) \dr \Go^2
\exp i[\lambda (\tau,\gb) u_\ga v^\ga +  (1-\tau) \gb v_\ga(y^\ga +p_+^\ga) +\Go^0_\ga
(u^\ga +y^\ga +p_+^\ga)]\,,\nn
\eee
where
\be
\Go_\ga =  (1-\tau) \gb v_\ga +\Go^0_\ga\,,
\ee
\be
\label{sW0}
\Go^0_\ga:= \Go_\ga \big |_{\gb=0} = \tau z_\ga - (1-\tau) (p_\ga (\gs) +\rho(  y_\ga
+p_{+\ga} +u_\ga )) \,,
\ee
\be
\label{lambda}
\lambda (\tau,\gb) := 1-(1-\tau)\gb\,.
\ee
Note that
 $\lambda >0$  in the allowed domain of the variables $\tau$ and $\gb$:
 \be\label{tb}
 \tau\in[0,1]\q-\infty <\gb<1\quad \Longrightarrow\quad \lambda (\tau,\gb)>0\,.
 \ee

The term linear in $v_\ga$ in the exponent can be removed by a change of integration variables
\be
u_\ga \to \tilde u_\ga=   u_\ga  -
(1- \lambda^{-1} (\tau,\gb))
(y_\ga +p_{+\ga}) \,,
\ee
\be
d u_\ga \to d \tilde u_\ga = d  u_\ga +
 d( \lambda^{-1} (\tau,\gb))
(y_\ga +p_{+\ga}) \,.
\ee

This yields
\be
X=\int_{\tilde u{}^2 v^2 \tau\gs\gb\rho }\ls\mu(\tau,\gs,\gb,\rho,u,v ) \dr \Go^2
\exp i[ u_\ga v^\ga\lambda(\tau,\gb) +\Go^0_\ga
( u^\ga +\lambda^{-1}(\tau,\gb)(y^\ga +p_+^\ga))]
\ee
with
\be
%\label{sW0=}
\Go^0_\ga:=  \tau z_\ga - (1-\tau) \big (p_\ga (\gs) +\rho\big(
\lambda(\tau,\gb)^{-1} (y_\ga +p_{+\ga}) + u_\ga \big )\big) \,
\ee
or, upon the redefinition,
\be
 u_\ga = \lambda(\tau,\gb)^{-1} u'_\ga,
\ee
\be
X=\int_{u'{}^2 v^2\tau\gs\gb\rho }\ls\mu(\tau,\gs,\gb,\rho, u,v )
 \dr \Go^2
\exp i[u^\prime_\ga v^\ga + \lambda(\tau,\gb)^{-1}\Go^0_\ga
(u'^\ga +y^\ga +p_+^\ga)]
\ee
with
\be
%\label{sW0==}
\Go^0_\ga:=  \tau z_\ga - (1-\tau) \big(p_\ga (\gs) +
\rho \lambda(\tau,\gb)^{-1} (y_\ga +p_{+\ga} +u^\prime_\ga) \big) \,,
\ee
\be
d \tilde u_\ga \to \lambda^{-1} (\tau,\gb) d u^\prime_\ga +
 d (\lambda (\tau,\gb)^{-1})u^\prime_\ga  \,.
\ee
Finally, making the rescaling
\be
\label{resc}
z_\ga = z_\ga' \lambda(\tau,\gb)\q
\gs_i = \gs_i{}' \lambda(\tau,\gb)\q \rho = \rho' \lambda(\tau,\gb)\,,
\ee
  we recast $X$ into the form
\be
X= \int_{u'{}^2 v^2\tau\gs\gb\rho }
\mu(\tau,\lambda \gs',\gb,\lambda\rho', u,v )
\dr \Go^2
\exp i[u^\prime_\ga v^\ga + \Go_\ga^0{}'
(u'^\ga +y^\ga +p_+^\ga)]\,,
\ee
\be
\label{sW0pr}
\Go_\ga^0 {}':=  \tau z_\ga ' - (1-\tau) \big (p_\ga (\gs') +
\lambda(\tau,\gb)^{-1} \rho' (y_\ga +p_{+\ga} +u^\prime_\ga)\big )  \,.
\ee
%with
%\be
%\Go_\ga = \lambda(\tau,\gb) \Go_\ga^0 {}' +(1-\tau)\gb v_\ga\,.
%\ee
Equivalently, taking into account that the $\rho$-dependent term does not contribute to the
exponent,
\be
X=\int_{u'{}^2 v^2\tau\gs\gb\rho }
\mu(\tau,\lambda \gs',\gb,\lambda\rho', u,v )
\dr \Go^2
\exp i[u^\prime_\ga v^\ga + \Go_\ga^0{}''
(u'^\ga +y^\ga +p_+^\ga)]\,,
\ee
\be
\label{sW0pr1}
\Go_\ga^0 {}'':=  \tau z_\ga ' - (1-\tau) \big ( p_\ga (\gs')  +
\rho' (u'_\ga +y_\ga +p_{+\ga})\big ) \,.
\ee
Since
$\Go_\ga^0 {}''$ (\ref{sW0pr1}) has  the same form as
$\Go_\ga$ at $\gb=0$,
$\gb$ can only affect  the vertex via the preexponential factor,
$\dr \Go^2 \mu(\tau,\lambda \gs',\gb,\lambda\rho', u, v )$
with
\be
\Go_\ga = \lambda(\tau,\gb) \Go_\ga^0 {}'' +(1-\tau)\gb v_\ga -(1-\tau)(1-\lambda(\tau,\gb))
\rho' (y_\ga +p_{+\ga} +u^\prime_\ga)\,.
\ee

Let us stress that  the exterior algebra formalism underlying the proposed
approach allows us to make any
 changes of  variables  $\tau, \g_i,\rho,\gb$ and  $z^\ga$.
 This is different from the analysis
 of \cite{Didenko:2019xzz} where the rescaling analogous to (\ref{resc}) was performed with
 the factor of $1-\gb $ treated as a constant parameter
  rather than $1-(1-\tau)\gb$ as in this paper. These
 rescalings coincide at $\tau=0$ and hence give rise to the same physical vertices
  at $\tau=0$. However, the approach of this paper allows one to
 trace the $\gb$-dependence effects  at the intermediate steps as well.

\section{Differential homotopy  form of HS equations}
\label{hdeq}

A remarkable novel feature of the proposed approach is that it treats
democratically space-time coordinates $x^n$, auxiliary spinor variables $Z^A$,
 homotopy parameters $t^a$ and even coordinates of the non-commutative
 star-product space like $U^A$ and $V^A$. This allows us to rewrite the
field equations in the {\it differential homotopy} form stripping away the integrations
over homotopy parameters $t^a$ and non-commutative coordinates $U^A$, $V^A$,
 extending the differential in the
$Z^A$ space to that in the $Z^A,\, t^a$, $U^A$, $V^A$, where $t^a$ and $U^A$, $V^A$
 denote the whole set of the homotopy parameters and noncommutative coordinates, respectively.
 For brevity, the dependence on the noncommutative coordinates is implicit
 in the rest of this section.

Thus, in addition to the space-time differential
$
\dr_x:= dx^n \f{\p}{\p x^n}\,,
$
we introduce {\it total exterior differential} $\dr$ (\ref{td}).

\subsection{Differential homotopy HS equations}
The transition to the differential homotopy setup with space-time coordinates
being on equal footing with the homotopy parameters demands slight
modification of the form of the HS equations.
 The idea is that upon integration
over homotopy parameters one has to recover the original system (\ref{HS1})-(\ref{HS5}).
For that purpose we let the field $B$ contain the homotopy differentials
$dt^a$ and $Z^A$-differentials $\theta^A$ as well as
space-time differentials that enter via the one-form $\go$. Hence
we set
\be
\label{cB}
\B=\sum_{n\geq 0} B^{(n)}
\ee
with $B^{(n)}$ containing $n$ one-forms $\go$
$$
B^{(n)}=\tilde B^n \go^n\,.
$$
Analogously $W$ is extended by the components being space-time
differential forms of higher  degrees.

It should be stressed that the new added fields will enter under
the $\dr$ differential and hence will not contribute to the field
equations at $Z^A=\theta^A=0$ upon integration over the homotopy parameters.
Thus, the differential homotopy system that we introduce now is equivalent
to the original integral one \eqref{HS1}-\eqref{HS5}.

Namely, introducing notation
\be
\label{cW}
\qquad\qquad \theta^A Z_A +\W := W+S\,,
\ee
where the  term $\theta^A Z_A$ is explicitly  introduced to incorporate
 $\dr_Z$ into $\dr$,
 the differential homotopy system of HS equations takes the form
\be
\label{B}
(\dr_x -2i \dr) \B + [\W\,, \B]_* \cong 0\,,
\ee
\be
\label{dW}
(\dr_x -2i\dr) \W + \W * \W \cong i\B*(\eta \gga +\bar \eta \bar\gga)\,,
\ee
where the sign $\cong$ implies {\it weak equality} up to terms that do not contribute under
integration.

There are three types of weak terms, that appear in the practical analysis.

 Firstly, these are the terms in the space $\V^<$ with the number
of the  homotopy differentials $dt^a$ smaller than that of homotopy
parameters to be integrated.

Secondly, these are the terms in the space $\V^>$ with the form degree greater
than  the number of homotopy
parameters to be integrated. Formally, such terms vanish because their
form degree exceeds that of the volume form. However, it is useful to keep them
because they may become relevant at higher orders after the number of homotopy
parameters (\ie the dimension of the homotopy space) is increased within the Ansatz of Section \ref{Ansatz}.

Thirdly, these are $\dr$-exact terms in $\V^E$
\be
v=\dr \phi\,.
\ee
Note that this is equivalent to the
space of $\dr_t$-exact terms since the difference $\dr\phi -\dr_t \phi =\dr_Z \phi$ is  in $\V^<$
for $\phi\in \V^<$.

The $\go$-dependent part of $\B$ now contributes into the
$\dr_x$ sector of the equation on the zero-form $C$ that acquires
 the form
\be
\dr_x C + [\go\,, C]_* -2i \dr (\tilde B^1 \go ) +\ldots =0\,.
\ee
Here the additional term with $\tilde B^1$ enters either under the
exterior differential in the homotopy parameters sector, that does
not contribute upon  integration, or under
 $\dr_Z$ differential which terms must cancel out in the $Z$-independent
(\ie $\dr_Z$-cohomological) sector of the dynamical equations.
Thus the modification of the $\B$ sector (\ref{cB}) does not affect the form
of integrated equations. Nevertheless, it still plays an important role in the
analysis of HS equations within the differential homotopy approach, being
demanded by the formal compatibility of the equations (\ref{B}) and
(\ref{dW}).

\subsection{Bulk and boundary terms in the homotopy space}
In the practical analysis it is convenient to distinguish between the bulk and boundary
terms in the space of homotopy parameters. The bulk terms are those that are integrated
over the interior of the square formed by the homotopy parameters $0< \tau<1$ and
$0< \bar\tau<1$ that stand in front of $z_\ga$ or $\bar z_\dga$ in $\Omega$ (\ref{sW})
or its conjugate.
 The boundary terms are those that are
supported by the boundary of the square, \ie $\tau$ and/or $\bar \tau$
equals to 0 or 1 containing a factor of $\De(\tau)$ or
$\De(\tau-1)$ and/or $\De(\bar \tau)$ or
$\De(\bar \tau-1)$. Note that the cohomological contribution to the dynamical
HS equations is located at $\tau=\bar\tau=0$.

For the degenerate square with no homotopy parameters $\tau$ and $\bar\tau$
 the corresponding fields are
treated as bulk. These are the HS fields $\go(Y;K|x)$ and
$C(Y;K|x)$.

A useful form of the homotopy-differential equations is
\be
\label{BV}
[(\dr_x -2i \dr) \B + [\W\,, \B]_*]\Big |_{bulk} =\V_\B\,,
\ee
\be
\label{dWV}
[(\dr_x -2i\dr) \W + \W * \W]\Big |_{bulk} =\V_{\W}\,,
\ee
where $\V_\B$ and $\V_{\W}$ are homotopy boundary terms. For instance, the \rhs of
(\ref{dW}) is a boundary term because $\gga$ and $\bar\gga$ (\ref{klein})
are supported at $\tau=1$ and  $\bar\tau=1$, respectively.
In practice we set
\be
\label{VW}
\V_\W \cong i\B * (\eta\gga + \bar \eta \bar \gga) \q \V_\B \cong 0 \,.
\ee

Now we observe that
$\gga$ can be represented in the form
(\ref{fbeta}) with $p_{i\ga}=0$
\be
\label{mu1=}
\mu= (\dr (l(\tau))-\De(\tau))\De(\gb)\De(\rho)\,,
\ee
where the first term is $\dr$-exact, while the second one yields a weakly zero result
upon multiplication by $\dr\Omega^2$ (\ref{sW}) since
there are no $\gs_i$ in that case. Thus, the boundary terms with $\gga$ and $\bar\gga$
can be represented  in the manifestly $\dr$-exact form modulo weakly zero terms.

As a consequence of the compatibility conditions for the equations
(\ref{BV}), (\ref{dWV}),
\be
\label{dvw}
(\dr_x -2i \dr) \V_\W + [\W\,, \V_\W]_* =0\,,
\ee
\be
\label{dVB}
(\dr_x -2i\dr) \V_\B + [\W\,, \V_\B ]_* =[\V_\W\,,\B]_*\,.
\ee
Note that $\V_\W$ (\ref{VW}) obeys these conditions as a consequence of (\ref{BV}) and
since $\gga$ and $\bar\gga$ are central.

The gauge transformations, that leave invariant equations
(\ref{BV}), (\ref{dWV}), (\ref{dvw}) and (\ref{dVB}) are
\be
\delta \W = (\dr_x -2i\dr)\epsilon_\W +[\W \,, \epsilon_\W]_*  \q
\delta \B = [\B \,, \epsilon_\W]_*\,,
\ee
\be
\delta \V_\W = [\V_\W\,, \epsilon_\W]_* \q
\delta \V_\B = [\V_\B \,, \epsilon_\W]_*\,,
\ee
\be
\label{epb}
\delta \B = (\dr_x -2i\dr)\epsilon_\B +[\W \,, \epsilon_\B]_*\q \delta
\V_B=[\V_\W\,,\epsilon_\B]\,,
\ee
where the
gauge parameters $\epsilon_\W$ and $\epsilon_\B$ are associated
with the fields $\W$ and $\B$, respectively.
Note that the gauge transformations with the parameters $\epsilon_\B$ are not
present in the original integral formulation of \cite{Vasiliev:1992av} (\ref{HS1})-(\ref{HS5})
where the field $B$ is a zero-form. In the differential formulation of this paper with $\B$
being a differential form both in $\M$ and in space-time, the gauge symmetry
parameters  $\epsilon_\B$ are responsible for the addition of exact forms, that do not
affect the final integrated result.

\section{Lower-order computations}
\label{LOC}

In this section we illustrate the differential homotopy machinery by the lower-order
analysis of the HS equations. Most of the results obtained  in this section were  reached before
within the  shifted homotopy analysis of \cite{Didenko:2018fgx} with the only important
exception of the reconstruction of
the projectively-compact spin-local form of the HS equations with the minimal number
of space-time derivatives, obtained originally ``by hand"
 in \cite{Vasiliev:2016xui}. The aim of this section is to illustrate
how the proposed approach works in practice. The higher-order analysis within the
differential homotopy scheme will be presented elsewhere.
Note that in this section we will use the differential version (\ref{pd}) for the
variables $p_\ga$ and $r_\ga$.

\subsection{Vacuum}
As a vacuum solution to equations (\ref{BV}), (\ref{dWV}) we choose as usual
\cite{Vasiliev:1992av} (see \cite{Vasiliev:1999ba} for a review)
\be
\B_0=0\q \W_0 = \go(Y;K|x)\q \V_{0\B}=0\q \V_{0\W} =0\,,
\ee
where $\go(Y;K|x)$ describes a flat HS connection that verifies
\be
\label{dgo}
\dr_x \go(Y;K|x) + \go(Y;K|x)* \go(Y;K|x) = 0\,.
\ee
For spin-two fields $\go(Y;K|x)$ bilinear in $Y^A$
\big($\go(Y;K|x) = \go^{AB}(x)Y_A Y_B$\big) these equations describe $AdS_4$ geometry
in terms of the flat $sp(4)$ connection $\go^{AB}(x)$.

\subsection{First order}

\subsubsection{$\B_1$}

In the lowest order, $\B_1$ is a bulk field on the degenerate homotopy square.
 Since the latter has no boundary,
equation (\ref{BV}) on $\B_1$ yields
\be
\label{db}
\D_x \B_1 -2i \dr\B_1 = 0 \,,
\ee
where $\D_x$ is the covariant derivative in the adjoint representation of the
star-product algebra,
\be\label{dx}
\D_x := \dr_x +[\go(Y;K|x)\,,\ldots ]_*\,.
\ee

The part of equation (\ref{db}) of degree zero as a space-time form implies that the $dx$-independent
part of $\B_1$ is $\dr$-closed. The gauge transformations (\ref{epb}) allow one to gauge
away the $\dr$-exact part. Choosing $\B_1 = C(Y;K|x)$ in $\dr$ cohomology
$ H^0(\dr)$ the $dx$-dependent part of equation (\ref{db}) yields the standard form of the linearized HS equations
 \cite{Vasiliev:1988sa},
\be
\D_x C(Y;K|x)=0\,.
\ee

\subsubsection{$S_1$}
$S_1$ determines  the first-order corrections to HS equations.
It is a sum of the terms linear in $\eta$ and $\bar \eta$,
\be
S_1=S_1^\eta + S_1^{\bar\eta}\,.
\ee
Equations (\ref{dWV}), (\ref{VW}) yield in the lowest order
\be
\label{S1}
-2i \dr S_1 \cong iC * (\eta\gga + \bar \eta \bar \gga)\,.
\ee
To reproduce the standard  linearized
HS equations of \cite{Vasiliev:1988sa}, where the $\eta$ deformation terms
contain the zero-form $C(y,\bar y;K|x)$ at $y=0$, \ie $C(0,\bar y;K|x )$, $ S^\eta_1$ has to be free from the
terms $y^\ga p_\ga$ in the exponent at $\tau=0$. (This is because the physical sector of the HS
equations is carried by $Z$-independent terms from $\dr_Z$ cohomology concentrated at $\tau=0$, which is
the content of the $Z$-dominance Lemma \cite{Didenko:2018fgx} (see also \cite{Didenko:2022eso}).)

Taking into account the form of $\Omega_\ga$ (\ref{sW})
\be
%\label{sW}
\Go_\ga (\tau,\gs,\gb,\rho) := \tau z_\ga - (1-\tau) (p_\ga \gs -\gb v_\ga +
\rho(  y_\ga +p_{\ga} +u_\ga )) \,
\ee
(note that $p_{+\ga}=p_\ga$ for the first order with a single $p_\ga$), we observe that
\be
\label{sW00}
\Go_\ga (\tau, \gs,\gb,\rho)\Big |_{\tau=0}
:= -  (p_\ga \gs- \gb v_\ga +\rho( y_\ga +p_\ga +u_\ga)) \,.
\ee
Hence,  the relevant part of the exponent at $\tau=0$ takes the form
\be
A=\exp i \big ( u_\ga v^\ga -  (p_\ga \gs- \gb v_\ga )(u^\ga+y^\ga +p^\ga)\big )\,.
\ee
The condition that it is free from $p_\ga y^\ga$  demands $\gs=0$.

Thus, we have to look for $S^\eta_1$ in the form (\ref{fbeta})
\be
\label{su1}
S_1^\eta
= -\f{\eta}{2} \int_{\tau\rho\gb\gs u^2 v^2}\ls d^2 u d^2 v\,
l(\tau)\mu(\gs,\rho,\gb) \dr \Go^\ga \dr \Go_\ga
\Ex C(r,\bar y;K) \big |_{r=0}k\,
\ee
with $l(\tau)$ (\ref{lt}) and
\be\label{mtri}
\mu (\gs,\rho,\gb) = \De(\gs)\tilde \mu (\rho,\gb) \,.
\ee
Analogously,
\be
\label{barsu1}
S_1^{\bar\eta}
= -\f{\bar \eta}{2} \int_{ \bar\tau\bar\rho\bar\gb\bar\gs \bar u^2\bar v^2}\ls d^2 \bar u d^2 \bar v\,
l(\bar\tau)\bar\mu(\bar\gs,\bar\rho,\bar \gb) \dr \bar \Go^\dga \dr \bar  \Go_\dga
\Exb C(y,\bar r;K) \big |_{\bar r=0}\bar k\,
\ee
with
\be
\bar \mu (\bar \gs,\bar \rho,\bar \gb) = \De(\bar\gs)\tilde {\bar \mu} (\bar\rho,\bar\gb) \,.
\ee

$S_1$ has to obey equation (\ref{S1}). Since $\gga$ and $\bar\gga$
have the form (\ref{klein}) this is possible if $\dr$ hits only $l(\tau)$ and the term with $\De(\tau)$
is weak. This demands $\tilde \mu (\rho,\gb)$ and $\tilde {\bar \mu} (\bar\rho,\bar\gb) $ be $\dr$-closed,
\be
\dr \tilde \mu (\rho,d\rho,\gb,d\gb )=0\q \dr\tilde {\bar \mu} (\bar\rho,d\bar\rho,\bar\gb,
d\bar\gb)=0\,,
\ee
which is not a restriction since
 any two-form holomorphic  in $\rho $ and $\gb$ or antiholomorphic in
 $\bar \rho $ and $\bar \gb$ is $\dr$-closed.

As a result, we can write
\be
\label{dsu1}
-2i\dr S_1^\eta
= i\eta  \int_{\tau\rho\gb\gs u^2 v^2}\ls\ls  d u^2 d v^2
(\De (1-\tau) +\De (\tau))\mu(\gs,\rho,\gb) \dr \Go^\ga \dr \Go_\ga
\Ex   C(r,\bar y;K) \big |_{r=0}k\,,
\ee
where the term with $\De(\tau)$ is weakly zero. Indeed, consider
\be
\V_{S_1^\eta} :=
 i\eta  \int_{\tau\rho\gb\gs u^2 v^2}\ls\ls du^2 dv^2
\De (\tau)\De(\gs) \tilde\mu(\rho,\gb) \dr \Go_0^\ga \dr \Go_{0\ga}
\exp i \big ( u_\ga v^\ga +\Go_{0\gb}(u^\gb+y^\gb +p^\gb)
\big) C(r,\bar y;K) \big |_{r=0}k\,,
\ee
where
\be
\Go_{0\ga} := \gb v_\ga -\rho(y_\ga +u_\ga +p_{\ga})\,.
\ee
In this case $\dr \Go_0^\ga \dr \Go_{0\ga}$ is a two-form
in $\gb$ and $\rho$ which, together with the two-form $\tilde\mu(\rho,\gb)$,
gives a four-form in the two-dimensional space, hence being weakly zero.
Though one might think that the weak terms have to
be discarded,  it is useful to keep them since
they  induce non-zero contribution at further computations respecting the form
of the Ansatz (\ref{fbeta}). In particular,  as explained in Section
\ref{LFE1},  the weak term $\V_{S_1^\eta}$    eventually determines the form of
 the linearized field equations.

The  term with $\De (1-\tau) $ yields
\be
%\label{dsu1}
-2i\dr S_1^\eta
\cong i\eta\int_{u^2 v^2} d^2u d^2 v  \int_{\rho\gb}
\tilde \mu(\rho,\gb)  \theta^\ga \theta_\ga
\exp i \big ( u_\ga v^\ga +z_\gb(u^\gb+y^\gb +p^\gb)
\big) C(r,\bar y;K) \big |_{r=0}k\,,
\ee
and, upon integration over $v$ and $u$,
\be
\label{ds1}
-2i\dr S_1^\eta
\cong i\eta\ \int_{\rho\gb}\
\tilde \mu(\rho,\gb) \theta^\ga \theta_\ga
\exp i \big ( z_\gb (y^\gb +p^\gb)
\big) C(r,\bar y;K) \big |_{r=0}k\,.
\ee
At the condition that
$
 \int_{\rho\gb}\
\tilde \mu(\rho,\gb)=1
$
this indeed gives $i\eta C* \gga $
(as a simple check, one can see that, at $C=1$ implying $p_\ga =0$,
(\ref{ds1}) reproduces $\gga$ (\ref{klein})).

To summarise, $S_1$ and $\V_{S_1}$, that solve (\ref{dWV})
within the differential homotopy approach in a way compatible with the
standard form of the free unfolded HS equations of \cite{Vasiliev:1988sa,Vasiliev:1999ba}, read as
\be
\label{s1f}
S_1^\eta
= -\f{\eta}{2} \int_{\tau\rho\gb\gs u^2 v^2}\ls \ls du^2 dv^2
l(\tau)\De(\gs)\tilde \mu(\rho,\gb) \dr \Go^\ga \dr \Go_\ga \Ex
 C(r,\bar y;K) \big |_{r=0}k\,,
\ee
\be
\label{vs1}
\V_{S_1^\eta} :=
 i\eta  \int_{\tau\rho\gb\gs u^2 v^2}\ls \ls du^2 dv^2
\De (\tau)\De(\gs) \tilde\mu(\rho,\gb) \dr \Go^\ga \dr \Go_{\ga}\Ex
C(r,\bar y;K) \big |_{r=0}k\,
\ee
with
\be
\label{sW1}
\Go_\ga  := \tau z_\ga - (1-\tau) (p_\ga \gs -\gb v_\ga +\rho(  y_\ga +p_{\ga} +u_\ga )) \,.
\ee

The conjugated expressions are
\be
\label{s1fb}
S_1^{\bar\eta}
= -\f{\bar\eta}{2} \int_{\bar \tau\bar\rho\bar\gb\bar\gs\bar u^2 \bar v^2}\ls\ls d \bar u^2
d\bar v^2
l(\bar \tau)\De(\bar \gs)\tilde{ \bar \mu}(\bar \rho,\bar \gb) \dr \bar \Go^\dga \dr\bar \Go_\dga
\Exb
C(y,\bar r;K) \big |_{\bar r=0}\bar k\,,
\ee
\be
\label{vs1b}
\V_{S_1^{\bar\eta}} :=
 i\bar \eta   \int_{\bar \tau\bar \rho\bar \gb\bar \gs \bar u^2 \bar v^2}\ls\ls d \bar u^2
d\bar v^2
\De (\bar \tau)\De(\bar \gs) \bar{\tilde\mu}(\bar \rho,\bar \gb) \dr \bar \Go^\dga \dr \bar
\Go_{\dga}\Exb
 C(y,\bar r;K) \big |_{\bar r=0}\bar k\,
\ee
with
\be
\label{sW1b}
\bar \Go_\dga := \bar \tau \bar z_\dga -
(1-\bar \tau) (\bar p_\dga \bar \gs -\bar \gb \bar v_\dga +
\bar \rho( \bar  y_\dga +\bar p_{\dga} +\bar u_\dga )) \,.
\ee
The closed two-forms $ \mu(\rho,\gb)$ and $\bar  \mu(\bar \rho,\bar \gb)$ are normalized to obey
\be
 \int_{\rho\gb}\
\tilde \mu(\rho,\gb)=1\q  \int_{\bar \rho\bar \gb}
\bar{\tilde \mu}(\bar \rho,\bar \gb)=1\,,
\ee
being otherwise arbitrary.

The conventional  solution for $S_1^\eta$ used  originally in \cite{Vasiliev:1992av}
(for review see \cite{Vasiliev:1999ba}) is a particular case
with
\be
\label{conmu}
\tilde \mu(\rho,\gb) = \De(\gb)\De (\rho)\,
\ee
implying zero $y-$ and $\f{\p}{\p y}-$ shifts in terms of the shifted homotopy of
\cite{Didenko:2018fgx}.

\subsubsection{$W_1$}

To find $W_1$, we have to solve the equation
\be
\label{dW1}
-2i \dr W_1 +\D_x S_1 = \V_{dW_1}\q
%D_0 := \dr_x +[\go\,,\ldots ]_*\,
\ee
with $\D_x$ (\ref{dx}).
$W_1$ contains four $\eta,\bar\eta$-dependent pieces of all possible kinds:
\be
W_1=\go(Y) + W_{1}^{\eta}|_{\go C} + W_1^{\eta}|_{C \go}+
W_{1}^{\bar \eta}|_{ \go C} + W_{1}^{\bar \eta}|_{ C \go}\,,
\ee
where the subscripts $\go C$ and $C\go$ refer to the order of the product factors
 while the superscripts $\eta$ and $\bar \eta$ distinguish between the terms
proportional to $\eta$ and $\bar\eta$, resp. Note that, to simplify formulae,
 we use convention of \cite{Vasiliev:1988sa} where $\omega$
contains both zero-order vacuum part and the first-order fluctuational part. Beyond the
vacuum approximation it is not demanded to obey (\ref{dgo}), however.

Using that $\dr_x C=-\{\go\,, C\}_* +\ldots,$ where ellipses denotes higher-order corrections, to evaluate $\D_x S_1$ it is convenient to use formulae (\ref{leftc}), (\ref{rightc})
which, taking into account that the one-form $\go$ anticommutes with the differentials of
the homotopy coordinates and, hence, with the  three-form $\mu(\gs,\gb,\rho )$ (\ref{mtri}),
yield
\be
\D_x S_1^{\eta}\big |_{\go C} =\f{\eta}{2}
\int_{\tau\gs \rho\gb\gs_\go }\ls\ls\dr P(\gb-1,\gs_\go,\gs) l(\tau)
\mu(\gs,\rho,\gb ) \dr \Go^2
\Ex
 \go(r_\go,\bar y )\bar * C(r_C,\bar y) \big  |_{r_{\go,C}=0} k \,,
\ee
\be
\ls\D_x S_1^{\eta}\big |_{C \go } =\f{\eta}{2}
\int_{\tau\gs\rho\gb\gs_\go }\ls\ls\dr P(\gs,\gs_\go,1-\gb) l(\tau)
\mu(\gs,\rho,\gb ) \dr \Go^2
\Ex
 C(r_C,\bar y) \bar * \go(r_\go,\bar y )\big  |_{r_{\go,C}=0}k \,.
\ee
(We use notations ($\bar *$) $\underline *$ for the star product in the (anti)holomorphic
sector, respectively. The integration variables $u^\ga$, $v^\ga$, $\bar u^\dga$ and $\bar v^\dga$
are implicit in the sequel.)
This yields
\be
\D_x S_1^{\eta}\big |_{\go C} =-\f{\eta}{2}\dr
\int_{\tau\gs\rho\gb\gs_\go }\ls\ls P(\gb-1,\gs_\go,\gs) l(\tau)\De(\gs)\tilde\mu(\rho,\gb)
\dr \Go^2
\Ex
 \go(r_\go,\bar y )\bar * C(r_C,\bar y) |_{r_{\go,C}=0} k +\V_{W_1^{\eta}|_{\go C}} \,,
\ee
\be
\D_x S_1^{\eta}\big |_{C \go } =-\f{\eta}{2}\dr
\int_{\tau\gs \rho\gb\gs_\go }\ls\ls P(\gs,\gs_\go,1-\gb)l(\tau) \De(\gs)
\tilde \mu(\rho,\gb ) \dr \Go^2
\Ex
 C(r_C,\bar y) \bar * \go(r_\go,\bar y ) |_{r_{\go,C}=0} k +\V_{W_1^{\eta}|_{ C\go}} \,,
\ee
where
\be
\ls\V_{W_1^{\eta}|_{\go C} } =-\frac{\eta}{2}
\int_{\tau\gs \rho\gb\gs_\go }\ls\ls P(\gb-1,\gs_\go,0) \dr(l(\tau))\De(\gs)\tilde\mu(\rho,\gb)
\dr \Go^2
\Ex
 \go(r_\go,\bar y )\bar * C(r_C,\bar y) |_{r_{\go,C}=0} k\,,
\ee
\be
\ls\V_{W_1^{\eta}|_{C \go }  }=-\frac{\eta}{2}
\int_{\tau\gs \rho\gb\gs_\go }\ls\ls P(0,\gs_\go,1-\gb) \dr(l(\tau))\De(\gs)\tilde\mu(\rho,\gb)
\dr \Go^2
\Ex
 C(r_C,\bar y) \bar *  \go(r_\go,\bar y )  |_{r_{\go,C}=0} k \,.
\ee

Here $\V_{W_1^{\eta}|_{\go C}} $ and $\V_{W_1^{\eta}|_{ C \go} }$ are weakly zero
boundary terms,
\be
\V_{W_1^{\eta}|_{\go C}}\cong0\q \V_{W_1^{\eta}|_{C\go}}\cong 0\,.
\ee
Indeed, since $\dr(l(\tau))=\De(\tau) +\De(1-\tau)$,
 the term with $\De(\tau)$,
 \be
 \De(\tau) \dr \Go_\ga = -\De(\tau)  (p_{\ga} d\gs + p_{\go\ga} d\gs_\go
  -d\gb v_\ga +d\rho(  y_\ga +p_{+\ga} +u_\ga ))
\ee
is a one-form in $d\gs,d\gs_\go, d \gb, d\rho$. As a result, the measure
factor is a five-form in these four differentials, hence being zero. Other way around,
the term
$
\De(1-\tau) \dr \Go_\ga = \De(1-\tau)\theta_\ga
$
does not contain the differentials $d\gs,d\gs_\go, d \gb, d\rho$ at all. Hence, in that case,
 the measure is a three-form in four variables that does not contribute to the integral
 as well.

This analysis illustrates  the important point that whether an expression is non-trivial
or weakly zero depends on how many homotopy parameters like $d\gs,d\gs_\go, d \gb, d\rho$
enter. Since some of them enter implicitly via  $\dr \Go_\ga$,
an expression can be weakly zero in presence of, say, $\gs_\omega$, being nontrivial at the
previous step without $\go$,  or other way around. This is why we prefer to
keep track of  all boundary terms no matter whether they are weakly zero or not.

From (\ref{dW1}) then follows
\be
\label{Wgoceta}
W_1^{\eta}|_{\go C} =\f{i\eta}{4}
\int_{\tau\gs\rho\gb \gs_\go }\ls P(\gb-1,\gs_\go,\gs) l(\tau)\De(\gs)\tilde\mu(\rho,\gb)
\dr \Go^2
\Ex
 \go(r_\go,\bar y )\bar * C(r_C,\bar y) \big  |_{r_{\go,C}=0} k \,,
\ee
\be
\label{Wcgoeta}
W_1^{\eta}|_{C \go } =\f{i\eta}{4}
\int_{\tau\gs \rho \gb \gs_\go }\ls P(\gs,\gs_\go,1-\gb)l(\tau) \De(\gs)
\tilde \mu(\rho,\gb ) \dr \Go^2
\Ex
 C(r_C,\bar y) \bar * \go(r_\go,\bar y )\big  |_{r_{\go,C}=0} k\,.
\ee

Analogously, in the antiholomorphic sector,
\be
W_{1}^{\bar \eta}|_{\go C} =\f{i\bar \eta}{4}
\int_{\bar\tau\bar\gs \bar\rho\bar\gb\bar\gs_\go }\ls P(\bar\gb-1,\bar\gs_\go,\bar\gs)
l(\bar\tau)\De(\bar\gs)\tilde{\bar \mu}(\bar\rho,\bar\gb)
\dr \bar \Go^2
\Exb
 \go(y,{\bar r}_\go ) \underline * C(y,{\bar r}_C) \big  |_{{\bar r}_{\go,C}=0} \bar k \,,
\ee
\be
W_{1}^{\bar \eta}|_{C \go } =\f{i\bar\eta}{4}
\int_{\bar\tau\bar\gs \bar\rho \bar\gb\bar{\gs}_\go }\ls\ls P(\bar\gs,\bar{\gs}_\go,1-\bar\gb)
l(\bar \tau) \De(\bar \gs)
\tilde {\bar \mu}(\bar\rho,\bar\gb ) \dr \bar\Go^2
\Exb
 C(y,{\bar r}_C)  \underline{*} \go(y,{\bar r}_\go,\bar y )\big  |_{{\bar r}_{\go,C}=0}\bar k \,
\ee
and
\be
\ls\V_{W_{1}^{\bar \eta}|_{\go C}} =-\frac{\bar\eta}{2}
\int_{\bar\tau\bar\gs \bar \rho \bar\gb{\bar \gs}_\go }\ls\ls P(\bar\gb-1,{\bar \gs}_\go,0)
\dr(l(\bar \tau))\De(\bar \gs)\tilde{\bar \mu}(\bar \rho,\bar \gb)
\dr \bar\Go^2
\Exb
 \go(y,{\bar r}_\go ) \underline{*} C(y,{\bar r}_C)\big  |_{{\bar r}_{\go,C}=0} \bar k\,,
\ee
\be
\ls\V_{W_{1}^{\bar \eta}|_{C\go}} =-\frac{\bar\eta}{2}
\int_{\bar \tau\bar \gs \bar \rho \bar \gb{\bar \gs}_\go }\ls \ls P(0,{\bar\gs}_\go,1-\bar\gb)
\dr(l(\bar\tau))\De(\bar \gs)\tilde{\bar \mu}(\bar\rho,\bar\gb)
\dr \bar\Go^2
\Exb
  C(y,{\bar r}_C)\underline{*}\go(y, {\bar r}_\go )  \big  |_{{\bar r}_{\go,C}=0} \bar k \,.
\ee

Another useful representation for $W_1$ is
\be\label{UWgoc}
W_1|_{\go C} = \frac{1}{2i} \Big (\int_{\tau\gs\rho\gb \gs_\go }\ls\ls
 P(\gb-1,\gs_\go,\gs)
\go(r,\bar y)\Big |_{r =0}\bar* S_1^{\eta} +
\int_{\bar\tau\bar\gs \bar\rho \bar\gb\bar\gs_\go }\ls\ls P(\bar\gb-1,\bar\gs_\go,\bar\gs)
\go (y,\bar r)\Big |_{\bar r=0}  \underline{*}
S_1^{\bar\eta} \Big )\,
\ee
with $S_1^\eta(\gs)$ (\ref{su1})
and $S_1^{\bar\eta} (\bar\gs)$ (\ref{barsu1}). Analogously,
\be
\V_{dW_1|_{\go C}} =-
\int_{\tau\gs\rho \gb \gs_\go }\ls\ls
 P(\gb-1,\gs_\go,\gs)
\go(r,\bar y)\Big |_{r =0} \bar * \dr S_1^{\eta} -
\int_{\bar\tau\bar\gs \bar\rho \bar\gb\bar\gs_\go }\ls\ls P(\bar\gb-1,\bar\gs_\go,\bar\gs)
\go (y,\bar r)\Big |_{\bar r=0}  \underline{*}
\dr S_1^{{\bar\eta}} \,
\ee
and
\be
\label{UWcgo}
W_{1} |_{C \go } =\f{1}{2i}\Big (
\int_{\tau\gs \rho \gb\gs_\go }\ls\ls P(\gs,\gs_\go,1-\gb)
S_1^{\eta} \bar
  * \go(r,\bar y )\Big   |_{r=0} +
\int_{\bar\tau\bar\gs \bar\rho \bar\gb\bar{\gs}_\go }\ls \ls P(\bar\gs,\bar{\gs}_\go,1-\bar\gb)
S_{1}^{\bar \eta}  \underline{*} \go(y,{\bar r} )\Big  |_{{\bar r}=0} \Big )
  \,,
\ee
\be
\V_{\dr W_1|_{C \go }} =
\int_{\tau\gs\rho\gb \gs_\go }\ls\ls P(\gs,\gs_\go,1-\gb)
\dr S_1^{\eta} \bar
  * \go(r,\bar y )\Big   |_{r=0} +
\int_{\bar\tau\bar\gs \bar\rho \bar\gb\bar{\gs}_\go }\ls\ls P(\bar\gs,\bar{\gs}_\go,1-\bar\gb)
\dr S_{1}^{\bar \eta}  \underline{*} \go(y,{\bar r} )\Big  |_{{\bar r}=0}
  \,.
\ee

\subsubsection{Linearised field equations in the one-form sector}
\label{LFE1}

The linearised equation on $\go$, that results from  (\ref{dWV}), yields
 \be
\label{D0}
\D_x \go + \D_x (W_1) +\dr W_{1,2} \cong \V_{\D_x \go}\,,
\ee
where $W_{1,2}$ is a space-time two-form and $\V_{\D_x \go}$ is weakly zero.
Consider the holomorphic sector.
%\subsubsection{Sector $\go\go C$}
Using (\ref{left}), (\ref{right}), (\ref{lr}) and  (\ref{l1l}), it is easy
to obtain
\be
\D_x W_1 \big |_{\go\go C}=
\f{\eta}{4i}
\int_{\tau\gs \rho\gb\gs_{\go_1}\gs_{\go_2} }\ls\ls\ls\ls \dr [P_3(\gb-1,\gs_{\go_1},
\gs_{\go_2},\gs)] l(\tau)\De(\gs)\tilde\mu(\rho,\gb)
\dr \Go^2
\Ex
 \go(r_{\go_1},\bar y )\bar *\go(r_{\go_2},\bar y )\bar * C(r_C,\bar y)\big  |_{r_{\go_i,C}=0}
  k \,.
\ee
Here the three terms in $\dr P_3$ correspond to $\go * W_1  |_{\go C}$,
$ W_1   |_{(\go*\go) C}$ and $ W_1   |_{\go (\go*C)}$.

Analogously,
\be
\D_x W_1 \big |_{C\go\go } =
\f{\eta}{4i}
\int_{\tau\gs \rho \gb \gs_{\go_1}\gs_{\go_2} }\ls\ls\ls\ls \dr [P_3(\gs,\gs_{\go_1},
\gs_{\go_2},1-\gb)] l(\tau)\De(\gs)\tilde\mu(\rho,\gb)
\dr \Go^2
\Ex
C(r_C,\bar y)\bar * \go(r_{\go_1},\bar y )\bar *\go(r_{\go_2},\bar y )\big  |_{r_{\go_i,C}=0}
  k \,.
\ee

The contribution to the $\go C\go$ sector consists of four terms
\bee
\D_x W_1^\eta |_{\go C\go}=&&\ls \f{i\eta}{4}\int_{\tau\gs\rho\gb\gs_{\go_1}\gs_{\go_2}}
\ls\ls\ls\big [ P_2(\gb-1,\gs_{\go_1},\gs)
\De(\gs_{\go_2} +\gb-1)-\De(\gs_{\go_1} +1-\gb)P_2(\gs, \gs_{\go_2},1-\gb)\nn\\&&\nn
\quad -\De (\gs_{\go_2} - \gs) P_2(\gb-1,\gs_{\go_1}, \gs) - \De(\gs - \gs_{\go_1})
P_2(\gs,\gs_{\go_2},1-\gb)\big ]\\&&
\label{W11}
\quad\times l(\tau)\De(\gs)\tilde\mu(\rho,\gb) \dr \Go^2
\Ex
\go(r_{\go_1},\bar y ) \bar * C(r_C,\bar y)\bar *\go(r_{\go_2},\bar y )
\big  |_{r_{\go_i,C}=0}  k \,,
\eee
where
the first two result, respectively, from $W_1|_{\go C} *\go$ and
$\go * W_1 |_{\,C \go }$ while the last two come from  $W_1|_{\,\go\dr_x  C}$
and $W_1|_{\dr_x C \go }$
with $\dr_x C =-[\go\,,C]_*+\ldots$ where ellipses denotes higher-order
terms that do not contribute to the order in question. This yields
\bee
\label{W12}
\D_x W_1^\eta   |_{\go C\go}=&&\ls \f{\eta}{4i}\int_{\tau\gs\rho\gb\gs_{\go_1}\gs_{\go_2}}
\ls\ls\ls\dr \big [ P_4(\gb-1,\gs_{\go_1},\gs,\gs_{\go_2},1-\gb) \big ]
\\&&\nn
\times l(\tau)\De(\gs)\tilde\mu(\rho,\gb)
\dr \Go^2
\Ex
\go(v_{\go_1},\bar y ) \bar * C(v_C,\bar y)\bar *\go(v_{\go_2},\bar y )
\big  |_{r_{\go_i,C}=0}  k \,.
\eee
To see that the expressions (\ref{W11}) and (\ref{W12}) are indeed equivalent,
one has to take into account inequalities that follow from the fact that $\gs=0$
due to the factor of $\De(\gs)$ and $1-\gb > 0$, that trivialize some of the
theta-function factors like, for instance,
$
\theta(\gs +1-\gb) = 1\,.
$
Note that such inequalities allow us to extend the $P_3$ factors in the $\go^2 C$ and
$C\go^2$ sectors to $P_4$ as follows:
\be
\label{K341}
P_3(\gb-1,\gs_{\go_1},\gs_{\go_2},\gs)\to P_4(\gb-1,\gs_{\go_1},
\gs_{\go_2},\gs, 1-\gb)\,,
\ee
\be
\label{K342}
P_3(\gs, \gs_{\go_1},
\gs_{\go_2}, 1-\gb)\to P_4(\gb-1,\gs, \gs_{\go_1},
\gs_{\go_2},1-\gb)\,.
\ee

This makes it possible  to write $\D_x W_1 $ in the form
\bee
\label{W1}
\D_x W^\eta_1 =&&\ls \f{i\eta}{4}\int_{\tau\gs\rho\gb\gs_{\go_1}\gs_{\go_2}}
\ \ls \ls\ls l(\tau)\De(\gs)\tilde\mu(\rho,\gb)
\dr \Go^2
\Ex\\&&
\Big [\dr \big [ P_4(\gb-1,\gs_{\go_1},\gs_{\go_2},\gs,1-\gb) \big ]
\go(r_{\go_1},\bar y ) \bar * \go(r_{\go_2},\bar y )\bar *C(r_C,\bar y)
\nn\\&&\nn
\ls + \dr\big [ P_4(\gb-1,\gs,\gs_{\go_1},\gs_{\go_2},1-\gb) \big ]
 C(r_C,\bar y)\bar *\go(r_{\go_1},\bar y ) \bar *\go(r_{\go_2},\bar y )
\\&&\nn
\ls +\dr \big [ P_4(\gb-1,\gs_{\go_1},\gs,\gs_{\go_2},1-\gb) \big ]
\go(r_{\go_1},\bar y ) \bar * C(r_C,\bar y)\bar *\go(r_{\go_2},\bar y )
\big ] \Big ] \big  |_{r_{\go_i,C}=0} k\,.
\eee
 Using that $\tilde\mu(\rho,\gb)$ is $\dr$-closed this allows us to find
$W_{1,2}^{\eta}$ modulo terms containing $\dr(l(\tau))=\De(\tau) +\De (1-\tau)$. The terms with
$\De (1-\tau)$ are weakly zero while those with $\De(\tau)$ give the contribution to the \rhs of
the field equations.
All in all this yields
\bee
\label{W121}
W_{1,2}^{\eta} =&&\ls \f{i\eta}{4}\int_{\tau\gs\rho\gb\gs_{\go_1}\gs_{\go_2}}
\ l(\tau)\De(\gs)\tilde\mu(\rho,\gb)
\dr \Go^2
\Ex\\&&
\Big [ P_4(\gb-1,\gs_{\go_1},\gs_{\go_2},\gs,1-\gb)
\go(r_{\go_1},\bar y ) \bar * \go(r_{\go_2},\bar y )\bar *C(r_C,\bar y)
\nn\\&&\nn
\ls + P_4(\gb-1,\gs,\gs_{\go_1},\gs_{\go_2},1-\gb)
 C(r_C,\bar y)\bar *\go(r_{\go_1},\bar y ) \bar *\go(r_{\go_2},\bar y )
\\&&\nn
\ls + P_4(\gb-1,\gs_{\go_1},\gs,\gs_{\go_2},1-\gb)
\go(r_{\go_1},\bar y ) \bar * C(r_C,\bar y)\bar *\go(r_{\go_2},\bar y )
\big ] \Big ] \Big  |_{r_{\go_i,C}=0} k\,,
\eee
\bee
\label{Vdr}
\V_{\D_x \go}^{\eta } =&&\ls \f{i\eta}{4}\int_{\tau\gs\rho\gb\gs_{\go_1}\gs_{\go_2}}
\De(1-\tau)\De(\gs)\tilde\mu(\rho,\gb)
\dr \Go^2 \Ex\\&&
\Big [ P_4(\gb-1,\gs_{\go_1},\gs_{\go_2},\gs,1-\gb)
\go(r_{\go_1},\bar y ) \bar * \go(r_{\go_2},\bar y )\bar *C(r_C,\bar y)
\nn\\&&\nn
\ls + P_4(\gb-1,\gs,\gs_{\go_1},\gs_{\go_2},1-\gb)
C(r_C,\bar y)\bar *\go(r_{\go_1},\bar y ) \bar *\go(r_{\go_2},\bar y )
\\&&\nn
\ls + P_4(\gb-1,\gs_{\go_1},\gs,\gs_{\go_2},1-\gb)
\go(r_{\go_1},\bar y ) \bar * C(r_C,\bar y)\bar *\go(r_{\go_2},\bar y )
\big ] \Big ] \Big  |_{r_{\go_i }=r_{ C}=0} k\,
\eee
in the holomorphic sector and
\bee
\label{W122}
W_{1,2}^{\bar\eta} =&&\ls \f{i\bar\eta}{4}\int_{\bar\tau\bar\gs\bar\rho \bar\gb\bar\gs_{\go_1}
\bar\gs_{\go_2}}
 l(\bar \tau)\De(\bar \gs)\tilde{\bar\mu}(\bar \rho,\bar \gb)
\dr \bar \Go^2
\Exb\\&&
\Big [ P_4(\bar \gb-1,\bar \gs_{\go_1},\bar \gs_{\go_2},\bar\gs,1-\bar\gb)
\go(y,\bar r_{\go_1} )  \underline{*} \go(y, \bar r_{\go_2} ) \underline{*}C(y,\bar r_C)
\nn\\&&\nn
\ls + P_4(\bar \gb-1,\bar\gs,\bar\gs_{\go_1},\bar\gs_{\go_2},1-\bar\gb)
  \underline{*} C(y,\bar r_C)\underline{*}\go(y,\bar r_{\go_1})  \underline{*}\go(y,\bar r_{\go_2} )
\\&&\nn
\ls + P_4(\bar\gb-1,\bar\gs_{\go_1},\bar\gs,\bar\gs_{\go_2},1-\bar\gb)
\go(y,\bar r_{\go_1})  \underline{*} C(y,\bar r_C) \underline{*}\go(y,\bar r_{\go_2} )
\big ] \Big ] \Big  |_{{\bar r}_{\go_i,C}=0} \bar k\,,
\eee
\bee
\label{Vdr=}
\V_{\D_x \go}^{\bar \eta } =&&\ls \f{i\bar \eta}{4}\int_{\bar \tau\bar \gs
\bar \rho \bar \gb\bar \gs_{\bar \go_1}\bar \gs_{\bar \go_2}}
\De(1-\bar \tau)\De(\bar \gs)\tilde{\bar \mu}(\bar \rho,\bar \gb)
\dr \bar \Go^2
\Exb\\&&
\Big [ P_4(\bar \gb-1,\bar \gs_{\go_1},\bar \gs_{\go_2},\bar \gs,1-\bar \gb)
\go(y,\bar r_{\go_1}) \underline{*} \go(y,\bar r_{\go_2}) \underline{*}C(y, \bar r_C, y)
\nn\\&&\nn
\ls + P_4(\bar \gb-1,\bar \gs,\bar \gs_{\go_1},\bar \gs_{\go_2},1-\bar\gb)
  C(y,\bar r_C) \underline{*}\go(y,\bar r_{\go_1} )  \underline{*}\go(y, \bar r_{\go_2})
\\&&\nn
\ls + P_4(\bar \gb-1,\bar \gs_{\go_1},\bar \gs,\bar \gs_{\go_2},1-\bar \gb)
\go(y, \bar r_{\go_1})  \underline{*} C(y, \bar r_C) \underline{*}\go(y, \bar r_{\go_2})
\big ] \Big ]\Big  |_{{\bar r}_{\go_i,C}=0}\bar k\,
\eee
in the antiholomorphic.

The field equations on the HS gauge fields are
\bee
\nn
\D_x \go =&&\ls \f{\eta}{4i}\int_{\tau\gs\rho\gb\gs_{\go_1}\gs_{\go_2}}
\De(\tau)\De(\gs)\tilde\mu(\rho,\gb)
\dr \Go^2
\Ex\\&&
\Big [ P_4(\gb-1,\gs_{\go_1},\gs_{\go_2},\gs,1-\gb)
\go(r_{\go_1},\bar y ) \bar * \go(r_{\go_2},\bar y )\bar *C(r_C,\bar y)
\nn\\&&\nn
\ls + P_4(\gb-1,\gs,\gs_{\go_1},\gs_{\go_2},1-\gb)
\bar * C(r_C,\bar y)\bar *\go(r_{\go_1},\bar y ) \bar *\go(r_{\go_2},\bar y )
\\&&\nn
\ls + P_4(\gb-1,\gs_{\go_1},\gs,\gs_{\go_2},1-\gb)
\go(r_{\go_1},\bar y ) \bar * C(r_C,\bar y)\bar *\go(r_{\go_2},\bar y )
\big ] \Big ]  \Big  |_{r_{\go_i,C}=0} k\nn
\\&&\label{dxgomu}
\ls +  \f{\bar \eta}{4i}\int_{\bar \tau\bar \gs\bar \rho \bar \gb\bar \gs_{\go_1}\bar \gs_{\go_2}}
\De(\bar \tau)\De(\bar \gs)\tilde{\bar \mu}(\bar \rho,\bar \gb)
\dr \bar \Go^2
\Exb\\&&
\Big [ P_4(\bar \gb-1,\bar \gs_{\go_1},\bar \gs_{\go_2},\bar\gs,1-\bar\gb)
\go(y, \bar r_{\go_1})  \underline{*} \go(y, \bar r_{\go_2})\underline{*}C(y, \bar r_C)
\nn\\&&\nn
\ls + P_4(\bar \gb-1,\bar \gs,\bar \gs_{\go_1},\bar \gs_{\go_2},1-\bar \gb)
 C(y, \bar r_C)\underline{*}\go(y, \bar r_{\go_1})  \underline{*}\go(y, \bar r_{\go_2})
\\&&\nn
\ls + P_4(\bar \gb-1,\bar \gs_{\go_1},\bar \gs,\bar \gs_{\go_2},1-\bar \gb)
\go(y, \bar r_{\go_1}) \underline{*} C(y, \bar r_C)\underline{*}\go(y, \bar r_{\go_2})
\big ] \Big ] \Big  |_{{\bar r}_{\go_i,C}=0} \bar k\,.
\eee

The \rhs of (\ref{dxgomu}) is in $\dr_Z$ cohomology because of the factors of $\De(\tau)$
or $\De (\bar \tau)$.
It is nonzero (not weak) since the pre-exponential factor is a six-form integrated over
a six-dimensional  homotopy space. On the other hand,
$\V_{\D_x \go}^{\eta}$ and $\V_{\D_x \go }^{\bar \eta}$ are  weak because  the factors
of $\De(1-\tau)$ or $\De(1-\bar\tau)$  eliminate
the differentials of homotopy coordinates from  $\dr\Go^2$ or $\dr\bar \Go^2$ leaving  a
four-form  integrated over the six-dimensional homotopy space, which is zero.

Equation (\ref{dxgomu}) reproduces the  standard form of Central On-Shell Theorem
at $\tilde \mu(\rho,\gb)$ (\ref{conmu}).
Indeed, from the fundamental Ansatz (\ref{fbeta}), (\ref{sW}), (\ref{Exp}) and the
fact that (\ref{dxgomu}) contains $\De(\gs)$ in the $\eta$ sector and
$\De(\bar\gs)$ in the $\bar \eta$ sector it follows that the \rhs contains $C(0,\bar y;K|x) $
and $C(y,0;K|x)$ in these sectors, respectively. This is the characteristic feature of the standard form of the Central On-Shell Theorem that fixes it uniquely by formal consistency.
Further details demand explicit evaluation of the various types of terms in (\ref{dxgomu})
analogous to the standard computation within the conventional formalism (see e.g.
\cite{Vasiliev:1999ba} and references therein).

 A related question is what is the effect of
 the specific choice of $\tilde \mu(\rho,\gb)$? The answer is that exact $\tilde \mu(\rho,\gb)$
do not contribute upon  integration. In the case of $\rho$, this is easy to see. Indeed,
since the two-form  $\tilde \mu(\rho,\gb)$ contains $\dr\rho$, this is the only place where
$\dr \rho$ contributes. As a result, neither $\dr \rho$ nor $\rho$ appears anywhere else
and, hence, the terms $\dr \rho \f{\p}{\p \rho} \phi(\rho,\gb)$ give rise
to the exact integrand in (\ref{dxgomu}).
It turns out that the $\gb$ dependence of $\tilde \mu(\rho,\gb)$ also does not contribute
at the lowest order which is consistent with the analysis of
Section \ref{gb}.

Another question is to which extent the form of the Central On-Shell Theorem can be affected
by the application of the general homotopy of Section \ref{General}. It is not hard
to make sure that the application of the general homotopy does not affect the field equations
except that it can add an exact form  not contributing upon integration over homotopy
parameters. So far, in the analysis of this section, the generalization of Section \ref{General}
could only affect the form of $S_1$ and $W_1$. As far as we can see, such a modification would
affect the form of the linearized HS equations (\ref{dxgomu}) that is not allowed as spoiling the  meaning of the components of the zero-form $C(Y;K|x)$ in terms of derivatives of the
HS gauge fields $\go(Y;K|x)$.

\subsection{Second order}
\subsubsection{$B_2$}
\label{B2}

The second-order part of $B$,
\be
B_2= B_2^\eta +B_2^{\bar\eta}\,,
\ee
 is determined by the equations
\be
\label{dBu}
2i\dr B_2^{\eta} = [S_1^{\eta}\,, C]_*-\V_{dB_2}^\eta\q
2i\dr B_2^{\bar\eta} = [S_1^{\bar\eta}\,, C]_*-\V_{dB_2}^{\bar\eta}
\ee
with $S_1^\eta$ (\ref{su1}) and $S_1^{\bar \eta}$ (\ref{barsu1}).
An elementary computation using (\ref{left}) and (\ref{right})  yields
\be
\ls S_1^\eta * C =
 -\f{\eta}{2} \int_{\tau\rho\gb\gs_1\gs_2}\ls\ls\De(\gs_2 - (1-\gb)) \De(\gs_1)
l(\tau)\tilde \mu(\rho,\gb) \dr \Go^2
\Ex C(r_1,\bar y;K) \bar * C(r_2,\bar y;K) \big  |_{{r_i}=0} k\,,
\ee
\be
C* S_1^\eta  =
 \f{\eta}{2} \int_{\tau\rho\gb\gs_1\gs_2}\ls\ls\De(\gs_1 + (1-\gb)) \De(\gs_2)
l(\tau)\tilde \mu(\rho,\gb) \dr \Go^2
\Ex C(r_1,\bar y;K) \bar * C(r_2,\bar y;K) \big  |_{{r_i}=0} k\,.
\ee
As a result,
\be
\label{s1c}
[S_1^\eta \,, C]_* =
 \f{\eta}{2} \int_{\tau\rho\gb\gs_1\gs_2}\ls\ls\De(\gs_2-\gs_1 - (1-\gb)) ( \De(\gs_2)-\De (\gs_1))
l(\tau)\tilde \mu(\rho,\gb) \dr \Go^2
\Ex C(r_1,\bar y;K) \bar * C(r_2,\bar y;K) \big  |_{{r_i}=0} k\,.
\ee
This expression can be represented in several different forms with the first two  insensitive
to the $\rho$-dependence:
\be
\label{sccon}
[S_1^\eta \,, C]_* =
 \f{\eta}{2} \int_{\tau\rho\gb\gs_1\gs_2}\ls\ls\dr [\theta(\gs_2-\gs_1 - (1-\gb)) ( \De(\gs_2)-\De (\gs_1))]
l(\tau)\tilde \mu(\rho,\gb) \dr \Go^2
\Ex C(r_1,\bar y;K) \bar * C(r_2,\bar y;K)\big  |_{{r_i}=0} k\,,
\ee
\be
\label{scshift}
[S_1^\eta \,, C]_* =
 -\f{\eta}{2} \int_{\tau\rho\gb\gs_1\gs_2}\ls\ls\dr [\De(\gs_2-\gs_1 - (1-\gb)) \theta(\gs_2)
 \theta (-\gs_1)]
l(\tau)\tilde \mu(\rho,\gb) \dr \Go^2
\Ex C(r_1,\bar y;K) \bar * C(r_2,\bar y;K) \big  |_{{r_i}=0} k\,.
\ee

Solving (\ref{dBu}) with the aid of (\ref{sccon}) and (\ref{scshift})
 yields, respectively, the following  results:
\be
\label{b2con}
B_{2 con}^\eta =
 \f{i\eta}{4} \int_{\tau\rho\gb\gs_1\gs_2}\ls\ls\theta(\gs_2-\gs_1 - (1-\gb)) ( \De(\gs_2)-\De (\gs_1))
l(\tau)\tilde \mu(\rho,\gb) \dr \Go^2
\Ex C(r_1,\bar y;K) \bar * C(r_2,\bar y;K)\big  |_{{r_i}=0} k\,,
\ee
\be
\label{b2shift}
B_{2sh}^\eta =
 \f{\eta}{4i} \int_{\tau\rho\gb\gs_1\gs_2}\ls\ls\De(\gs_2-\gs_1 - (1-\gb))  \theta(\gs_2)
 \theta (-\gs_1) l(\tau)\tilde \mu(\rho,\gb) \dr \Go^2
\Ex C(r_1,\bar y;K) \bar * C(r_2,\bar y;K) \big  |_{{r_i}=0} k\,
\ee
and the weak terms
\be
\label{v2con}
\V_{B_{2 con}}^\eta =-
 \f{\eta}{2} \int_{\tau\rho\gb\gs_1\gs_2}\ls\ls\dr (l(\tau))\theta(\gs_2-\gs_1 - (1-\gb)) ( \De(\gs_2)-\De (\gs_1))
\tilde \mu(\rho,\gb) \dr \Go^2
\Ex C(r_1,\bar y;K) \bar * C(r_2,\bar y;K)\big  |_{{r_i}=0} k\,,
\ee
\be
\label{v2shift}
\V_{B_{2sh}}^\eta =
 \f{\eta}{2} \int_{\tau\rho\gb\gs_1\gs_2}\ls\ls\dr (l(\tau))\De(\gs_2-\gs_1 - (1-\gb))  \theta(\gs_2)
 \theta (-\gs_1) \tilde \mu(\rho,\gb) \dr \Go^2
\Ex C(r_1,\bar y;K) \bar * C(r_2,\bar y;K) \big  |_{{r_i}=0} k\,.
\ee

$B_2{}_{con}$ and $B_2{}_{sh}$ can be easily recognized to reproduce
those resulting, respectively, from the conventional (unshifted) homotopy, that
leads to the non-local field equations,  and the shifted homotopy
of \cite{Gelfond:2018vmi}, that leads to the spin-local HS
equations. It should be stressed that spin-locality of the vertex associated with
$B_{2sh}^\eta$ is due to  the factor of $\De(\gs_2-\gs_1 - (1-\gb)) $ on the \rhs
of (\ref{b2shift}) that guarantees the cancellation of the terms with $p_{1\ga} p_2^\ga$ in
the exponent at $\tau=0$. (For more detail see Section \ref{EdC}.)

However, being spin-local, the resulting vertex is not
projectively-compact \cite{Vasiliev:2022med} which means that it is not of the lowest order in derivatives and
most important,  can induce space-time non-local  terms at higher-orders. Though the proper projectively-compact
HS vertex was  found in \cite{Vasiliev:2016xui}  by hand
field redefinition, so far it was not known how to reach this result by
 a systematic homotopy method. One of the results of this paper is  the
elaboration of such a method, that is  based on the extension of the approach by the
parameter $\rho$.

To this end, setting
\be
\tilde \mu (\rho,\gb)=-\De(\rho) \tilde \mu (\gb)\q \int_\gb \tilde \mu(\gb)=1\,
\ee
in (\ref{s1c}) yields
\bee
[S_1^\eta \,, C]_* &&\ls
=\f{\eta}{2} \int_{\tau\rho\gb\gs_1\gs_2}\ls\ls\De(\gs_2-\gs_1 - (1-\gb)) ( \De(\gs_1+\rho)-
\De (\gs_2 -\rho))%\label{1-2}
\nn\\&&
l(\tau)\De(\rho)\tilde \mu(\gb) \dr \Go^2
\Ex C(r_1,\bar y;K) \bar * C(r_2,\bar y;K) \big  |_{{r_i}=0} k
\,,
\eee
where
\be
\Go_\ga  := \tau z_\ga - (1-\tau) (p_\ga(\gs) -\gb v_\ga +\rho( p_{+\ga} +u_\ga
+ y_\ga) )\q p_\ga(\gs)= \gs_1 p_{1\ga} +\gs_2 p_{2\ga}  \,.
\ee
Now, we make a change of variables of the class (\ref{frr}) in
the term with $\De(\gs_2-\rho)$
\be
\label{rho}
\rho\to \rho' = 1-\gb -\rho\,.
\ee
This yields
\bee
[S_1^\eta \,, C]_* &&\ls
=\f{\eta}{2} \Big (\int_{\tau\rho\gb\gs_1\gs_2}\ls\ls  \De(\gs_1+\rho)
\De(\rho)(\dr \Go(\rho))^2 + \int_{\tau\rho\gb\gs_1\gs_2}\ls\ls
\De (\gs_2 +\rho-(1-\gb))\De (1-\gb -\rho)(\dr \Go(\rho'))^2\Big )\nn\\&&\De(\gs_2-\gs_1 - (1-\gb)
)
\label{1-2}
l(\tau)\tilde \mu(\gb)\Ex C(v_1,\bar y;K) \bar * C(v_2,\bar y;K) k
\,,
\eee
where it is used that $\Ex$ is
$\rho$-independent and $\int_{1-\gb-\rho}=-\int_{\rho}$.

Taking into account that the measure factors in (\ref{1-2}) contain $d\rho'$, we observe that
\be
\De(1-\gb-\rho) (\dr \Go (1-\gb-\rho))^2 = \De(1-\gb-\rho)
\Big [ (\dr \Go(\rho))^2
-2 (1-\gb)d\tau (y^\ga +u^\ga +p^\ga_+)\dr\Go_\ga(\rho)\Big ] \,
\ee
and, hence, using relations analogous to (\ref{sw+}), (\ref{w}), (\ref{dOm}) and (\ref{fbetawd}),
\bee
[S_1^\eta \,, C]_* = &&\ls-\f{\eta}{2} \int_{\tau\rho\gb\gs_1\gs_2}\ls\ls l(\tau)
\dr\Big(\De
(\gs_1 +\rho ) \De (\gs_2 -(1-\gb -\rho))
\big (\theta(\rho) \theta(1-\gb -\rho) \dr \Go^2 \label{noo}\\&&
+2i(1-\gb)\De (1-\gb-\rho) d\tau\big)%\nn\\&&
\tilde \mu(\gb)\Ex \Big ) C(r_1,\bar y;K)  \bar * C(r_2,\bar y;K)\big  |_{{r_i}=0} k
\nn
\,.
\eee
 Then (\ref{dBu}) gives $B_{2pc}^\eta$ that leads to a projectively-compact
 spin-local vertex,
 \bee
 \label{B2prc}
B_{2pc}^\eta = &&\ls\f{\eta}{4i} \int_{\tau\rho\gb\gs_1\gs_2}\ls\ls l(\tau)
\De
(\gs_1 +\rho ) \De (\gs_2 -(1-\gb -\rho))
\big (\theta(-\gs_1) \theta(\gs_2) \dr \Go^2 +2i(1-\gb)\De (\gs_2) d\tau \big )\nn\\&&
\tilde \mu(\gb)\Ex C(r_1,\bar y;K) \bar * C(r_2,\bar y;K)\big  |_{{r_i}=0} k
\,,
\eee
\bee
\label{VdB2prc}
\V^\eta_{dB_{2pc}} = &&\ls\f{\eta}{2} \int_{\tau\rho\gb\gs_1\gs_2}\ls\ls \dr (l(\tau))
\De
(\gs_1 +\rho ) \De (\gs_2 -(1-\gb -\rho))
\theta(-\gs_1) \theta(\gs_2) \dr \Go^2 \nn\\&&
\tilde \mu(\gb)\Ex C(r_1,\bar y;K) \bar * C(r_2,\bar y;K)\big  |_{{r_i}=0} k
\,,
\eee
where the  vanishing term with $\dr (l(\tau)) d\tau$ has been omitted.
The term $\V^\eta_{dB_{2pc}}$ is weakly zero because $\De(\tau) \dr \Go^2$ and
 $\De(1-\tau) \dr \Go^2$ contain, respectively,  too many and not enough differentials versus
   the number of integration variables.

 The antiholomorphic expressions are
 \bee
 \label{B2prcb}
B_{2pc}^{\bar \eta} = &&\ls\f{\bar \eta}{4i} \int_{\bar\tau\bar\rho\bar\gb\bar\gs_1\bar\gs_2}
\ls\ls l(\bar\tau)
\De
(\bar\gs_1 +\bar \rho ) \De (\bar \gs_2 -(1-\bar \gb -\bar \rho))
\big(\theta(-\bar \gs_1) \theta(\bar \gs_2) \dr \bar {\Go}^2 +2i(1-\bar\gb)\De (\bar \gs_2) d\bar \tau)\big )\nn\\&&
\tilde{\bar  \mu}(\bar \gb)\Exb C( y, \bar r_1;K)   \underline{*} C (y,\bar r_2;K) \big  |_{{\bar r_i}=0} \bar k
\,,
\eee
        \bee
\label{VdB2prcb}
\V^{\bar \eta}_{dB_{2pc}} = &&\ls\f{\bar \eta}{2} \int_{\bar \tau\bar \rho\bar \gb
\bar \gs_1\bar \gs_2}\ls\ls \dr (l(\bar \tau))
\De
(\bar \gs_1 +\bar \rho ) \De (\bar \gs_2 -(1-\bar \gb -\bar \rho))
\theta(-\bar \gs_1) \theta(\bar \gs_2) \dr \bar {\Go}^2
\nn\\&&
\tilde {\bar \mu}(\bar \gb)\Exb C(y, \bar r_1;K) \underline{*} C(y,\bar r_2;K) \big  |_{{\bar r_i}=0} \bar k
\,.
\eee
We shall see in the next section that the nontrivial contribution to
the field equations results from the weak term (\ref{VdB2prc}). The key
property of $B_{2pc}^\eta$ is  that  the resulting vertices are
projectively-compact, that eventually both minimizes the number of space-time derivatives  and guarantees equivalence
 of the space-time and spinor
spin-locality of the equations at the higher order. As explained in Sections
\ref{EdC} and \ref{pc}, this follows from the specific form of $B_2$
containing the differentials $\dr\gs_i$ and
$\dr \rho$ either via their sum $\dr \gs_i + \dr \rho$ or via $\dr \Go^2$.

Note that the $\dr \Go^2$--independent term in (\ref{noo})  in our construction is a counterpart of the shift $\delta B_2$  found in
\cite{Vasiliev:2016xui} to reach the vertex with the minimal number of derivatives.%%

\subsubsection{Equations ${\mathcal D}C = J$}
\label{EdC}
The $\dr_xC$ part of equation (\ref{BV}) has the form
\be
 \D_x C +[W_1\,, C]_*  +\D_x B_{2pc} -2i \dr B_{2 \go}= \V_{\D_x B_{2pc}} \,,
\ee
where $\V_{\D_x B_2}$ is the weak part that does not contribute upon
integration over homotopy parameters and
$\D_x C$ is the free part of the equations. With the aid of
(\ref{UWgoc}), (\ref{UWcgo}), (\ref{leftc})-(\ref{midl}) and (\ref{B2prc}) this yields in the holomorphic $\go CC$ sector
\bee
\label{dc}
\V^\eta_{\D_x B_{2pc}}\big |_{\go CC} =&&\ls \D_x C + \f{1}{2i} \int_{\gs_\go} P(\gb-1,\gs_\go,\gs_1) S_1^\eta (\go C_1) * C_2
+\dr \Big [\int_{\gs_\go} P(\gb-1,\gs_\go,\gs_1)B_{2pc}^\eta (\go  C_1 C_2 )\Big ]\nn \\&&
+\int_{\gs_\go} P(\gb-1,\gs_\go,\gs_1) \dr B_{2pc}^\eta(\go C_1 C_2)
 -2i \dr B^\eta_{2\go} \big |_{\go CC} \,,
\eee
where the arguments like $\go C_1$ or $\go  C_1 C_2 $ indicate the order of the
 product factors in the respective expressions resulting from insertion of the factors of $\go$
via formulae (\ref{leftc})-(\ref{midl}).
To cancel the $\dr$-exact term we set
\be
 B^{\eta}_{2\go}\big |_{\go CC} =\f{1}{2i} \int_{\gs_\go} P(\gb-1,\gs_\go,\gs_1)B_{2pc}^\eta (\go  C_1 C_2) \,.
\ee

Now we observe that
the first  term with $S^\eta_1 *C$ cancels against the analogous term from
$\dr B_{2pc}^\eta$ (\ref{dBu}). The term $C*S^\eta_1$ from $\dr B^\eta_{2pc}$ does not contribute because,
by (\ref{left}), it is proportional to $\De (\gb-1 -\gs_1)$ hence giving zero
since $P(\gb-1,\gs_0,\gb-1) =0$ in agreement with the property (\ref{Kk}) that
the arguments of $P(a_1,a_2\ldots )$ are ordered, $a_i\leq a_{i+1}$.
As a result, the only contributing term results
from the weak term (\ref{VdB2prc}) in $\dr B_{2pc}$ (\ref{dBu}) in
\be
\int_{\gs_\go} P(\gb-1,\gs_\go,\gs_1) \dr B_{2pc}^\eta(\go C_1 C_2)\,.
\ee
The term (\ref{VdB2prc}) consists of two parts. The one with $\De(1-\tau)$
is still weak determining $\V^\eta_{\D_x B_2|_{\go CC}} $,
\bee
\label{Vgocc2prc}
\V^\eta_{\D_x B_2|_{\go CC}}  = &&\ls\f{ i\eta }{4}
\int_{\tau\rho\gb\gs_\go\gs_1\gs_2}\ls\ls \De(1-\tau)P(\gb-1,\gs_\go,\gs_1)
\De
(\gs_1 +\rho ) \De (\gs_2 -(1-\gb -\rho))
\theta(-\gs_1) \theta(\gs_2) \dr \Go^2 \nn\\&&
\tilde \mu(\gb)\Ex \go(r_\go,\bar y;K)\bar *  C(r_1,\bar y;K) \bar * C(r_2,\bar y;K)\big |_{r_{\go, C_i}=0} k
\,.
\eee
That with $\De(\tau)$
is cohomological (\ie $Z$, $\theta$-independent).
It determines the nonlinear correction $J^\eta_{pc} \big |_{\go CC}$ to the HS equations,
\bee
\label{Jgocc}
J^{\eta}|_{\go CC}  = &&\ls\f{ i \eta }{4}
\int_{\tau\rho\gb\gs_\go\gs_1\gs_2}\ls\ls \De(\tau)P(\gb-1,\gs_\go,\gs_1)
\De
(\gs_1 +\rho ) \De (\gs_2 -(1-\gb -\rho))
\theta(-\gs_1) \theta(\gs_2) \dr \Go^2  \nn\\&&
\tilde \mu(\gb)\Ex\go(r_\go,\bar y;K)\bar *  C(r_1,\bar y;K) \bar * C(r_2,\bar y;K)\big |_{r_{\go, C_i}=0} k
\,.
\eee

Other two orderings are analysed analogously. The bulk terms in $\tau$ cancel
up to $\dr$-exact terms. (Note that in the sector $C\go C$
the bulk terms between $[W_1\,,C]_*$ and $P\dr B_2$ cancel pairwise.)
The final result has concise  symmetric form
\be
\D_x C= J^\eta_{pc} + J^{\bar \eta}_{pc}+ \V^{ \eta}_{\D_x B_2 }  + \V^{\bar \eta}_{\D_x B_2  }\,
\ee
with
\bee
\label{Jgo2ceta}
J^{\eta}_{pc}  = &&\ls\f{ i \eta }{4}
\int_{u^2 v^2 \tau\rho\gb\gs_\go \gs_1\gs_2}\ls\ls d^2 u d^2 v \De(\tau)
\De
(\gs_1 +\rho ) \De (\gs_2 -(1-\gb -\rho))
\theta(-\gs_1) \theta(\gs_2) \dr \Go^2 \tilde \mu(\gb)\Ex  \nn\\&&
\Big [ P(\gb-1,\gs_\go,\gs_1,\gs_2,1-\gb)\go(r_\go,\bar y;K)\bar *C(r_1,\bar y;K) \bar * C(r_2,\bar y;K)
\\&&\
+ P(\gb-1,\gs_1,\gs_\go,\gs_2,1-\gb) C(r_1,\bar y;K)\bar * \go(r_\go,\bar y;K) \bar * C(r_2,\bar y;K)\nn\\&&
+ P(\gb-1,\gs_1,\gs_2,\gs_\go,1-\gb)
 C(r_1,\bar y;K) \bar * C(r_2,\bar y;K)\bar * \go(r_\go,\bar y;K)\Big ] \Big |_{r_{\go, C_i}=0} k\nn
\,,
\eee
\bee
\label{Jgo2cbeta}
J^{\bar \eta}_{pc}  = &&\ls\f{  i \bar\eta }{4}
\int_{\bar u^2 \bar v^2 \bar \tau\bar \rho\bar \gb\bar \gs_\go \bar \gs_1\bar \gs_2}\ls\ls d^2 \bar u d^2 \bar v \De(\bar \tau)
\De
(\bar \gs_1 +\bar \rho ) \De (\bar \gs_2 -(1-\bar \gb -\bar \rho))
\theta(-\bar \gs_1) \theta(\bar \gs_2) \dr \bar \Go^2 \tilde {\bar \mu}(\gb)\Exb  \nn\\&&
\Big [ P(\bar \gb-1,\bar \gs_\go,\bar \gs_1,\bar \gs_2,1-\bar \gb )\go( y,\bar r_\go, ;K)
\underline{*}C( y,\bar r_1;K) \underline{*}
C(y, \bar r_2;K)
\\&&\
+ P(\bar \gb-1,\bar \gs_1,\bar \gs_\go,\bar \gs_2,1-\bar \gb) C(y, \bar r_1;K) \underline{*}
\go(y,\bar r_\go;K)  \underline{*} C(y,\bar r_2;K)\nn\\&&
+ P(\bar \gb-1,\bar \gs_1,\bar \gs_2,\bar \gs_\go, 1-\bar \gb) C(y,\bar r_1;K)  \underline{*} C(y,\bar r_2;K) \underline{*}
\go(y,\bar r_\go;K)\Big ] \Big |_{\bar r_{\go, C_i}=0}\bar k\nn
\,,
\eee
\bee
\label{Vdb2eta}
 \V^{ \eta}_{\D_x B_2  }  = &&\ls\f{i \eta  }{4 }
\int_{u^2 v^2 \tau\rho\gb\gs_\go\gs_1\gs_2}\ls\ls \De(1-\tau) d^2u d^2 v
\De
(\gs_1 +\rho ) \De (\gs_2 -(1-\gb -\rho))
\theta(-\gs_1) \theta(\gs_2) \dr \Go^2 \tilde \mu(\gb)\Ex  \nn\\&&
\Big [ P(\gb-1,\gs_\go,\gs_1,\gs_2,1-\gb)\go(r_\go,\bar y;K)\bar *C(r_1,\bar y;K) \bar * C(r_2,\bar y;K)
\\&&\
+ P(\gb-1,\gs_1,\gs_\go,\gs_2,1-\gb) C(r_1,\bar y;K)\bar * \go(r_\go,\bar y;K) \bar * C(r_2,\bar y;K)\nn\\&&
+ P(\gb-1,\gs_1,\gs_2,\gs_\go,1-\gb)
 C(r_1,\bar y;K) \bar * C(r_2,\bar y;K)\bar * \go(r_\go,\bar y;K)\Big ] \Big |_{r_{\go, C_i}=0} k\nn
\,,
\eee
\bee
\label{Vdb2beta}
\V^{\bar \eta}_{\D_x B_2  }  = &&\ls\f{ i \bar \eta }{4}
\int_{\bar u^2 \bar v^2\bar \tau\bar \rho\bar\gb\bar\gs_\go \bar \gs_1\bar \gs_2}\ls\ls
\De(1-\bar \tau) d^2 \bar u d^2 \bar v
\De
(\bar \gs_1 +\bar \rho ) \De (\bar \gs_2 -(1-\bar \gb -\bar \rho))
\theta(-\bar \gs_1) \theta(\bar \gs_2) \dr \bar \Go^2 \tilde {\bar \mu}(\gb)\Exb  \nn\\&&
\Big [ P(\bar \gb-1,\bar \gs_\go,\bar \gs_1,\bar \gs_2,1-\bar \gb )\go( y,\bar r_\go, ;K)
\underline{*}C( y,\bar r_1;K) \underline{*}
C(y, \bar r_2;K)
\\&&\
+ P(\bar \gb-1,\bar \gs_1,\bar \gs_\go,\bar \gs_2,1-\bar \gb) C(y, \bar r_1;K) \underline{*}
\go(y,\bar r_\go;K)  \underline{*} C(y,\bar r_2;K)\nn\\&&
+ P(\bar \gb-1,\bar \gs_1,\bar \gs_2,\bar \gs_\go, 1-\bar \gb) C(y,\bar r_1;K)  \underline{*} C(y,\bar r_2;K) \underline{*}
\go(y,\bar r_\go;K)\Big ]\big |_{r_{\go, C_i}=0} \Bar k\nn
\,,
\eee
with
\be
\Go^\ga =\tau z^\ga -(1-\tau)(\gs_\go p_\go^{\ga} +\gs_1 p_{C_1}^{\ga} +\gs_2 p_{C_2}^{\ga} -\gb v^\ga +
\rho(y^\ga + u^\ga+ p_+^{\ga}))
\q p_+^\ga := p_{\go}^\ga +p_{C_1}^\ga +p_{C_2}^\ga\,.
\ee

Note that the ordering of the arguments of the kernels $P$ in (\ref{Jgo2ceta}), (\ref{Jgo2cbeta}) matches that  of the
factors $\go$ and $C_{1,2}$. Though these parts of the formulae are symmetric with
respect to the exchange of $\go$ and $C_{1,2}$, the overall factor distinguishes
between $\go$ and $C_{1,2}$ in a way prescribed by the proper form of the Central
on-Shell Theorem and projectively-compact spin-locality of the currents $J$. Also note that some of the
$\theta$-functions in the definition of $P$ (\ref{Kk}) may trivialize
upon multiplication by the overall measure factor. For instance, one can see that the
first term in (\ref{Jgo2ceta}) reproduces  (\ref{Jgocc}).

The expressions for $\V_{\D_x B_2}^ \eta$ and
$\V_{\D_x B_2}^{\bar \eta} $ are weak (do not contribute)
since they contain only two differentials for the four
variables $\rho$,   $\gs_{1,2}$ and $\gs_\go$ because the factor of $\De (1-\tau)$
annihilates the part of $\dr \Go_\ga$ that contains $d\rho$,  $d\gs_{1,2}$ and $d\gs_\go$.
On the other hand, the currents $J^{ \eta}$
 are non-zero because in this case
$\dr \Go^2$  just brings two more differentials
in  $d\rho$,  $d\gs_{1,2}$ and $d\gs_\go$. (Analogously in the antiholomorphic sector.)

It should be stressed that the weak term in (\ref{dBu}) contributes
to the field equation for $C$ just because the latter contains an extra
homotopy parameter $\gs_\go$ associated with the one-from $\go$.
In other words, while in (\ref{dBu}) the measure contained a four-form
in the three-dimensional homotopy space, in the field equations
it is still a four-form which however is integrated over a four-dimensional
homotopy space with the extra coordinate $\gs_\go$.
 This phenomenon illustrates a general feature of  the proposed formalism making
 it useful to keep track of the weak terms that may
contribute at the later stage.

Let us now consider more in detail the structure of the current
$J^{ \eta}$ to show that it is spin-local.
(Recall that spin-locality \cite{Gelfond:2019tac} means that the restriction
of the vertex to any finite subset of fields is local. The vertex is not local in the
usual sense because  $\bar * (\underline{*})$ products in the  $\eta(\bar\eta)$ sector
 mix vertices of fields of different spins.)
 Since
locality is determined by the terms bilinear in $p_i$ in the exponent (\ref{Exp}),
it suffices  to keep there  only the $y$-independent terms at $\tau=0$:
\be
\int d^2 u d^2 v \exp i[-(p_{1\ga} \gs_1 +  p_{2\ga} \gs_2 -\gb v_\ga)(p_1^\ga +p_2^\ga+u^\ga) +
u_\ga v^\ga - p_{1\ga}p_2^\ga]\,.
\ee
Neglecting the $\gb$-dependent Jacobian, upon integration over $u_\ga$ and $v_\ga$
this yields
\be
 \exp i\big[\big (\f{\gs_2 -  \gs_1}{1-\gb} -1\big )
 p_{1\ga}p_2^\ga\big ]\,.
\ee
Since the preexponential factor $\De(\gs_1 + \rho ) \De ( \gs_2 -(1- \gb - \rho))$
in (\ref{Jgo2ceta}) sets $\gs_2 -\gs_1 = 1-\gb$, in agreement
with  \cite{Gelfond:2018vmi,Didenko:2020bxd}
this  implies the absence of  $p_{1\ga}p_2^\ga$  in the exponent and, hence,
spin-locality. That the constructed
 vertex is projectively-compact is explained in Section \ref{pc}.

\section{General lessons}
\label{less}
Let us highlight some general features of the developed formalism illustrated by the
analysis of the previous section.

\subsection{Cancellation of the bulk homotopy  terms}
\label{cbt}

As we have seen, the bulk terms in the homotopy space with $\tau\in (0,1)$
must cancel out in the final result leaving only the boundary terms located
at $\tau=\bar\tau =0$. This fact was known before as the $Z$-dominance Lemma
\cite{Didenko:2018fgx} (see also \cite{Didenko:2022eso}).) The novel feature of the
proposed formalism is that now the bulk homotopy terms cancel out pairwise without detailed
computations that in the previously available  formalisms demanded essential use
of the Schouten identity. This remarkable simplification is due to the
fundamental Ansatz (\ref{fbeta}), (\ref{sW}) that trivializes the role of the
Schouten identity.

Let us stress that the final result for the dynamical field equations must
belong to $\dr_Z$-cohomology in the original integral formulation of Section
\ref{HSsketch1}. Indeed, one can see that $\dr_Z$ of the \lhs
of equations (\ref{HS1}) and (\ref{HS3}) yields zero  once
equations (\ref{HS2}), (\ref{HS4}), (\ref{HS5}) have been solved.
This gives a perturbative proof that equations
 (\ref{HS1}) and (\ref{HS3}) are $Z$-independent.

 In the differential homotopy approach of this paper the analogous statement is that
 the \lhs of equations (\ref{B}) and (\ref{dW}) in the physical sector must be $Z,\theta$
 independent upon integration over the homotopy space $\M$, which in turn implies that such terms
 must be $\dr$-exact. These however are cancelled by the terms like $W_{1,2}$ of Section
 \ref{LFE1} or $B_{2\go}$ of Section \ref{EdC}. Thus all bulk terms in physical HS equations
 have to cancel out while the nontrivial contribution is concentrated at $\tau=\bar\tau=0$.
 (The terms at $\tau=1$ and/or $\bar\tau=1$ are weakly zero in physical equations.)

\subsection{Projectively-compact spin-locality}
\label{pc}

Generally, spin-locality in the spinor space of $Y$ variables does not automatically imply
space-time spin-locality at higher orders. As was explained  recently in \cite{Vasiliev:2022med}, there exists however
a class of projectively-compact spin-local vertices for which this is true.
The characteristic feature of such vertices, which, in fact, contain a minimal possible
number of derivatives \cite{Vasiliev:2016xui}, is that they contain contraction between
HS connection $\go(Y)$ and the $Y$-variables. In terms of vertices this
demands them  to contain the contraction $y_\ga p_\go^{ \ga}$ or $\by_\dga \bar p_\go^{ \dga}$.
The approach of this paper gives hints how to control the projective compactness
of the resulting spin-local vertices.
To illustrate this point let us show
that the vertices (\ref{Jgo2ceta}) do indeed share this property.

For simplicity,  consider the case of $\gb=0$  (\ie $\mu(\gb)=\De (\gb)$) where
the $u,v$ integration trivializes. (The analysis for general $\gb$
can be done analogously.)
We observe that the factor of $\De(\gs_1 + \rho ) \De ( \gs_2 -(1 - \rho))$
implies that the combinations of the differentials $\dr \gs_1 +\dr \rho$ and
$\dr \gs_2 + \dr \rho$ drops out from $\dr \Go^2$. As a result,
the terms from  $\dr \Go^2$ that survive at $\tau=0$ are
\be
\De (\tau) \De(\gs_1 + \rho ) \De ( \gs_2 -(1 - \rho)) \dr \Go^2 =
2 \De (\tau) \De(\gs_1 + \rho ) \De ( \gs_2 -(1 - \rho))\dr \rho \dr
\gs_{\go}y^\ga p_{\go\ga}\,,
\ee
which just  implies the projective compactness \cite{Vasiliev:2022med} of the resulting
vertex.
As explained in \cite{Vasiliev:2022med}, this guarantees equivalence of
space-time and spinor spin-locality in the analysis of the higher-order
interactions, that makes this choice distinguished
for the higher-order analysis.

  Thus,  the form of
$B_2$ (\ref{B2prc}) leads to the field equations containing an overall factor of
$y^\ga p_{\go\ga}$.  Our analysis shows that, more generally, to reach
projective compactness of the vertex of this type at higher orders
the prexponent factors have to be of the form
\be
\prod_{i}\De(\gs_{C_i} +\rho +\ldots)\,
\ee
in which case $d\rho$ and $d\gs_\go$ will contribute via $\dr \Go^2$ along with
the factor of  $y^\ga p_{\go\ga}$. This simple observation gives  a hint
what is a proper form of the general projectively-compact vertices in the
proposed formalism. It would be interesting to understand better its relation with the
shift symmetry condition of \cite{Didenko:2022eso}.

\section{Conclusion}
\label{con}

In this paper, a new approach to the analysis of HS equations of \cite{Vasiliev:1992av}
 is proposed with the homotopy parameters, that result from the Poincar\'e resolution of
 the differential equations,  treated democratically with  the space-time coordinates
 and non-commutative coordinates of the auxiliary spinor space. The geometric form of the
 approach is very suggestive: extra coordinates associated with the homotopy parameters
 parameterize compact polyhedra embedded into a multidimensional hypercube.
 From that perspective, they are reminiscent of the coordinates of the extra compact spaces like
 Calabi-Yau  in String Theory, {\it etc}. Since the original
 unfolded approach to HS theory of \cite{Vasiliev:1988sa, Vasiliev:1992av} is formulated
 entirely in terms of the exterior algebra machinery, the formulation is coordinate independent, allowing
 in particular to reformulate the theory in terms of smooth manifolds rather than polyhedra.

 The {\it differential homotopy}
 approach shares two seemingly conflicting features: it is both far more general and far simpler
 than the standard shifted homotopy approach of \cite{Didenko:2018fgx}. One of its great advantages is that it trivializes the complicated issue of accounting the Schouten identity. The same time,
 the proposed approach allowed us to derive HS current interactions in the projectively-compact
 spin-local form, that, as shown in \cite{Vasiliev:2022med}, implies equivalence of the spinor
 and space-time spin-locality. (Though this form of HS vertices was obtained by hand
 field redefinitions in \cite{Vasiliev:2016xui}, so far it was unreachable by systematic
 homotopy approaches.) Moreover, the differential homotopy approach will be used in \cite{OG} to derive
 moderately non-local $\eta\bar\eta$ vertices, establishing some bound on
 non-locality of  HS gauge theory. This raises two related interesting problems. One is
 to compare the level of non-locality of HS theory found in \cite{OG} with that derived by
 Sleight and Taronna \cite{Sleight:2017pcz} from the Klebanov-Polyakov holographic conjecture
 \cite{KP}. Another one is to
 look for better schemes that may further reduce the level of non-locality of HS theory at higher orders.

Another remarkable output  of the analysis of this paper is that, as explained in
Section \ref{cbt}, in the process of computation of the terms that contribute to field
equations most of them (so-called homotopy  bulk terms) cancel out by using general
HS equations without going into their detailed form.
To apply this mechanism to higher-order computations, where it is anticipated to be
particularly efficient, it is desirable to work out its extension to multilinear
combinations of the elementary fields. We hope to come back to this interesting
problem in the future.

 We believe that the results of this paper highlight the beauty and deepness of HS gauge theory
 to the extent making its study extremely interesting and exciting. The generality of the proposed approach gives a hope to uncover its relationship  with
 other powerful tools in field theory like, for instance, the harmonic superspace approach to supersymmetric  models \cite{Galperin:1984av}
 extended recently to  HS multiplets
 \cite{Buchbinder:2022svx}-\cite{Ivanov:2023uik}.

Let us stress that the differential homotopy approach
tested in this paper within the $AdS_4$ HS
theory is   applicable to other HS models like $3d$ HS gauge theory of \cite{Prokushkin:1998bq}
and HS gauge theory in any dimension of \cite{Vasiliev:2003ev}. The form of the
equations is still (\ref{B}), (\ref{dW})
 while the procedure is exactly
the same as in the $4d$ theory considered in this paper though the master fields
may depend on somewhat different auxiliary variables. Namely, one introduces the homotopy
space $\M$, strips out integrations and adds additional components of fields that contribute
under the total differential hence not affecting the final integrated results. Since the
extension to other HS models is straightforward we will not discuss it in more detail here.
Analogously, the differential homotopy approach can be applied to the multi-particle Coxeter HS theory \cite{Vasiliev:2018zer}  anticipated to shed light on the relation between
HS theory and String Theory,  one of the favorite  ideas of Lars Brink.

  \section*{Acknowledgments}
I am  grateful to Olga Gelfond for careful reading the manuscript and  many
useful discussions and comments. I  acknowledge with gratitude useful
comments by Vyatcheslav Didenko, Philipp Kirakosiants, Aleksandr Tarusov, Kirill Ushakov,
Danila Valeriev, and, especially,  Vyatcheslav Vereitin  and the referee. Also I wish to thank
Ofer Aharony, Theoretical High Energy Physics Group of Weizmann Institute of Science
for hospitality at the final stage of this work.

\addcontentsline{toc}{section}{\,\,\,\,\,\,\,References}

\section*{}


\begin{thebibliography}{99}
\parindent=0pt
\parskip=0pt

%\small

%\cite{Fronsdal:1978rb}
\bibitem{Fronsdal:1978rb}
C.~Fronsdal,
%``Massless Fields with Integer Spin,''
Phys. Rev. D
\textbf{18} (1978), 3624.

%\cite{Fradkin:1987ks}
\bibitem{Fradkin:1987ks}
  E.~S.~Fradkin and M.~A.~Vasiliev,
  %``On the Gravitational Interaction of Massless Higher Spin Fields,''
  Phys.\ Lett.\ B {\bf 189} (1987) 89.

%\cite{Fradkin:1986ka}
\bibitem{Fradkin:1986ka}
E.~S.~Fradkin and M.~A.~Vasiliev,
%``Candidate to the Role of Higher Spin Symmetry,''
Annals Phys. \textbf{177} (1987), 63.


%\cite{Berends:1984rq}
\bibitem{Berends:1984rq}
F.~A.~Berends, G.~J.~H.~Burgers and H.~van Dam,
%``On the Theoretical Problems in Constructing Interactions Involving Higher Spin Massless Particles,''
Nucl. Phys. B \textbf{260} (1985), 295-322.

%\cite{Vasiliev:1989qh}
\bibitem{Vasiliev:1989qh}
M.~A.~Vasiliev,
%``Quantization on sphere and high spin superalgebras,''
JETP Lett. \textbf{50} (1989), 374-377.



\bibitem{Bergshoeff:1989ns}
E.~Bergshoeff, M.~P.~Blencowe and K.~S.~Stelle,
%``Area Preserving Diffeomorphisms and Higher Spin Algebra,''
Commun. Math. Phys. \textbf{128} (1990), 213.
%doi:10.1007/BF02108779



%\cite{Vasiliev:1989re}
\bibitem{Vasiliev:1989re}
M.~A.~Vasiliev,
%``Higher Spin Algebras and Quantization on the Sphere and Hyperboloid,''
Int. J. Mod. Phys. A \textbf{6} (1991), 1115-1135.
%doi:10.1142/S0217751X91000605

%\cite{Prokushkin:1998bq}
\bibitem{Prokushkin:1998bq}
  S.~F.~Prokushkin and M.~A.~Vasiliev,
  %``Higher spin gauge interactions for massive matter fields in 3-D AdS space-time,''
  Nucl.\ Phys.\ B {\bf 545} (1999) 385
  [hep-th/9806236].



%\cite{Bengtsson:1983pd}
\bibitem{Bengtsson:1983pd}
  A. K. H. Bengtsson, I. Bengtsson and L. Brink,
  %``Cubic Interaction Terms for Arbitrary Spin,''
 {\it  Nucl.Phys.} {\bf B227} (1983) 31.

%\cite{Gelfond:2018vmi}
\bibitem{Gelfond:2018vmi}
O.~A.~Gelfond and M.~A.~Vasiliev,
%``Homotopy Operators and Locality Theorems in Higher-Spin Equations,''
Phys. Lett. B \textbf{786} (2018), 180-188
[arXiv:1805.11941 [hep-th]].



%\cite{Vasiliev:1987zv}
\bibitem{Vasiliev:1987zv}
  M.~A.~Vasiliev,
  %``Massless Fields Of All Spins In The Anti-de Sitter Space And Their Gravitational Interaction,''
  in *Sellin 1987, Proceedings, Theory of elementary particles* 234-252.

 %\cite{Gross:1987ar}
\bibitem{Gross:1987ar}
  D.~J.~Gross and P.~F.~Mende,
  %``String Theory Beyond the Planck Scale,''
  Nucl.\ Phys.\ B {\bf 303} (1988) 407.

%\cite{Gross:1988ue}
\bibitem{Gross:1988ue}
  D.~J.~Gross,
  %``High-Energy Symmetries of String Theory,''
  Phys.\ Rev.\ Lett.\  {\bf 60} (1988) 1229.


%\cite{Vasiliev:2018zer}
\bibitem{Vasiliev:2018zer}
M.~A.~Vasiliev,
%``From Coxeter Higher-Spin Theories to Strings and Tensor Models,''
JHEP \textbf{08} (2018), 051
%doi:10.1007/JHEP08(2018)051
[arXiv:1804.06520 [hep-th]].

%\cite{Brink:1992xr}
\bibitem{Brink:1992xr}
L.~Brink, T.~H.~Hansson and M.~A.~Vasiliev,
%``Explicit solution to the N body Calogero problem,''
Phys. Lett. B \textbf{286} (1992), 109-111
%doi:10.1016/0370-2693(92)90166-2
[arXiv:hep-th/9206049 [hep-th]].

%\cite{Brink:1993sz}
\bibitem{Brink:1993sz}
L.~Brink, T.~H.~Hansson, S.~Konstein and M.~A.~Vasiliev,
%``The Calogero model: Anyonic representation, fermionic extension and supersymmetry,''
Nucl. Phys. B \textbf{401} (1993), 591-612
%doi:10.1016/0550-3213(93)90315-G
[arXiv:hep-th/9302023 [hep-th]].



\bibitem{Maldacena:1997re}
J.~M.~Maldacena,
  %``The large N limit of superconformal field theories and supergravvity,''
  Adv.\ Theor.\ Math.\ Phys.\  {\bf 2} (1998) 231
  [Int.\ J.\ Theor.\ Phys.\  {\bf 38} (1999) 1113]
  [arXiv:hep-th/9711200].
  %%CITATION = IJTPB,38,1113;%%

\bibitem{Gubser:1998bc}
S.~S.~Gubser, I.~R.~Klebanov and A.~M.~Polyakov,
  %``Gauge theory correlators from non-critical string theory,''
  Phys.\ Lett.\  B {\bf 428}, 105 (1998)
  [arXiv:hep-th/9802109].
  %%CITATION = PHLTA,B428,105;%%

\bibitem{Witten:1998qj}
E.~Witten,
  %``Anti-de Sitter space and holography,''
  Adv.\ Theor.\ Math.\ Phys.\  {\bf 2}, 253 (1998)
  [arXiv:hep-th/9802150].
  %%CITATION = 00203,2,253;%%

\bibitem{KP} I.~R.~Klebanov and A.~M.~Polyakov,
  %``AdS dual of the critical O(N) vector model,''
  Phys.\ Lett.\ B {\bf 550} (2002) 213
    [hep-th/0210114].




\bibitem{Sezgin:2002rt}
  E.~Sezgin and P.~Sundell,
  %``Massless higher spins and holography,''
  Nucl.\ Phys.\  B {\bf 644} (2002) 303
  [Erratum-ibid.\  B {\bf 660} (2003) 403]
  [arXiv:hep-th/0205131].
  %%CITATION = NUPHA,B644,303;%%



\bibitem{LP}  R.~G.~Leigh and A.~C.~Petkou,
  %``Holography of the N=1 higher spin theory on AdS(4),''
  JHEP {\bf 0306} (2003) 011 [hep-th/0304217].

%\cite{Sezgin:2003pt}
\bibitem{Sezgin:2003pt}
E.~Sezgin and P.~Sundell, %``Holography in 4D (super) higher spin
%theories and a test via cubic scalar couplings,''
JHEP \textbf{07} (2005), 044
%doi:10.1088/1126-6708/2005/07/044
[arXiv:hep-th/0305040 [hep-th]].


%\cite{Aharony:2011jz}
\bibitem{Aharony:2011jz}
  O.~Aharony, G.~Gur-Ari and R.~Yacoby,
  %``d=3 Bosonic Vector Models Coupled to Chern-Simons Gauge Theories,''
  JHEP {\bf 1203} (2012) 037
  [arXiv:1110.4382 [hep-th]].
  %%CITATION = ARXIV:1110.4382;%%

%\cite{Giombi:2011kc}
\bibitem{Giombi:2011kc}
S.~Giombi, S.~Minwalla, S.~Prakash, S.~P.~Trivedi, S.~R.~Wadia and X.~Yin,
%``Chern-Simons Theory with Vector Fermion Matter,''
Eur. Phys. J. C \textbf{72} (2012), 2112
%doi:10.1140/epjc/s10052-012-2112-0
[arXiv:1110.4386 [hep-th]].

\bibitem{FF} M.~Flato and C.~Fronsdal, {Lett. Math. Phys.} {\bf 2}, 421 (1978);
{\it Phys. Lett.} B {\bf 97}, 236 (1980).



%\cite{Bekaert:2015tva}
\bibitem{Bekaert:2015tva}
  X.~Bekaert, J.~Erdmenger, D.~Ponomarev and C.~Sleight,
  %``Quartic AdS Interactions in Higher-Spin Gravity from Conformal Field Theory,''
  JHEP {\bf 1511} (2015) 149
 % doi:10.1007/JHEP11(2015)149
  [arXiv:1508.04292 [hep-th]].


%\cite{Sleight:2017pcz}
\bibitem{Sleight:2017pcz}
C.~Sleight and M.~Taronna,
%``Higher-Spin Gauge Theories and Bulk Locality,''
Phys. Rev. Lett. \textbf{121} (2018) no.17, 171604
%doi:10.1103/PhysRevLett.121.171604
[arXiv:1704.07859 [hep-th]].



%\cite{Ponomarev:2017qab}
\bibitem{Ponomarev:2017qab}
  D.~Ponomarev,
  %``A Note on (Non)-Locality in Holographic Higher Spin Theories,''
  Universe {\bf 4} (2018) no.1,  2
 % doi:10.3390/universe4010002
  [arXiv:1710.00403 [hep-th]].

%\cite{Neiman:2023orj}
\bibitem{Neiman:2023orj}
Y.~Neiman,
%``Quartic locality of higher-spin gravity in de Sitter and Euclidean anti-de Sitter space,''
[arXiv:2302.00852 [hep-th]].

%\cite{Bekaert:2005jf}
\bibitem{Bekaert:2005jf}
X.~Bekaert, N.~Boulanger and S.~Cnockaert,
% ``Spin three gauge theory revisited,''
JHEP \textbf{01} (2006), 052
%doi:10.1088/1126-6708/2006/01/052
[arXiv:hep-th/0508048 [hep-th]].
%68 citations counted in INSPIRE as of 10 Jan 2021

%\cite{Bengtsson:2006pw}
\bibitem{Bengtsson:2006pw}
A.~K.~H.~Bengtsson, %``Structure of higher spin gauge interactions,''
 J. Math. Phys. \textbf{48} (2007), 072302
[arXiv:hep-th/0611067 [hep-th]].


\bibitem{Manvelyan:2010jr}
R.~Manvelyan, K.~Mkrtchyan and W.~Ruhl, %``General trilinear
%interaction for arbitrary even higher spin gauge fields,''
 Nucl.\
Phys.\ B {\bf 836} (2010) 204 [arXiv:1003.2877].

%\bibitem{Manvelyan:2010je} R. Manvelyan, K. Mkrtchyan and W. Ruehl, {\it Phys.Lett.} {\bf B696} (2011) 410 [arXiv:1009.1054].

%\cite{Sagnotti:2010at}
\bibitem{Sagnotti:2010at}
A.~Sagnotti and M.~Taronna,
%``String Lessons for Higher-Spin Interactions,''
Nucl. Phys. B \textbf{842} (2011), 299-361
%doi:10.1016/j.nuclphysb.2010.08.019
[arXiv:1006.5242 [hep-th]].
%167 citations counted in INSPIRE as of 10 Jan 2021

%\cite{Fotopoulos:2010ay}
\bibitem{Fotopoulos:2010ay}
  A.~Fotopoulos and M.~Tsulaia,
  %``On the Tensionless Limit of String theory, Off - Shell Higher Spin Interaction Vertices and BCFW Recursion Relations,''
  JHEP {\bf 1011} (2010) 086
 % doi:10.1007/JHEP11(2010)086
  [arXiv:1009.0727 [hep-th]].

%\cite{Vasilev:2011xf}
\bibitem{Vasilev:2011xf}
  M.~A.~Vasiliev,
  %``Cubic Vertices for Symmetric Higher-Spin Gauge Fields in $(A)dS_d$,''
  Nucl.\ Phys.\ B {\bf 862} (2012) 341
 % doi:10.1016/j.nuclphysb.2012.04.012
  [arXiv:1108.5921 [hep-th]].

%\cite{Joung:2012rv}
\bibitem{Joung:2012rv}
  E. Joung, L. Lopez and M. Taronna,
  %``On the cubic interactions of massive and partially-massless higher spins in (A)dS,''
  {\it JHEP} {\bf 1207} (2012) 041
 % doi:10.1007/JHEP07(2012)041
  [arXiv:1203.6578].
  %%CITATION = doi:10.1007/JHEP07(2012)041;%%
  %53 citations counted in INSPIRE as of 16 Oct 2017


%\cite{Buchbinder:2012xa}
\bibitem{Buchbinder:2012xa}
I.~L.~Buchbinder, T.~V.~Snegirev and Y.~M.~Zinoviev, %``On
%gravitational interactions for massive higher spins in $AdS_3$,''
J. Phys. A \textbf{46} (2013), 214015
%doi:10.1088/1751-8113/46/21/214015
[arXiv:1208.0183 [hep-th]].

%\cite{Francia:2016weg}
\bibitem{Francia:2016weg}
D.~Francia, G.~L.~Monaco and K.~Mkrtchyan,
%``Cubic interactions of
%Maxwell-like higher spins,''
JHEP \textbf{04} (2017), 068
%doi:10.1007/JHEP04(2017)068
[arXiv:1611.00292 [hep-th]].

%\cite{Buchbinder:2017nuc}
\bibitem{Buchbinder:2017nuc}
I.~L.~Buchbinder, S.~J.~Gates and K.~Koutrolikos, %``Higher Spin
%Superfield interactions with the Chiral Supermultiplet: Conserved
%Supercurrents and Cubic Vertices,''
 Universe \textbf{4} (2018)
no.1, 6
%doi:10.3390/universe4010006
[arXiv:1708.06262 [hep-th]].
%31 citations counted in INSPIRE as of 10 Jan 2021

%\cite{Buchbinder:2023xlb}
\bibitem{Buchbinder:2023xlb}
I.~L.~Buchbinder and A.~A.~Reshetnyak,
%``Consistent Lagrangians for irreducible interacting higher-spin fields with holonomic constraints,''
[arXiv:2304.10358 [hep-th]].

%\cite{Bengtsson:2012jm}
\bibitem{Bengtsson:2012jm}
A.~K.~H.~Bengtsson, ``Systematics of Higher-spin Light-front
Interactions,'' [arXiv:1205.6117 [hep-th]].


%\cite{Metsaev:1991mt}
\bibitem{Metsaev:1991mt}
R.~R.~Metsaev, %``Poincare invariant dynamics of massless higher
%spins: Fourth order analysis on mass shell,''
Mod. Phys. Lett. A
\textbf{6} (1991), 359-367.
%doi:10.1142/S0217732391000348

\bibitem{Metsaev:1993} R. R. Metsaev, {\it Mod.Phys.Lett.} {\bf A8} (1993) 2413–-2426.

%\cite{Metsaev:1999ui}
\bibitem{Metsaev:1999ui}
R.~R.~Metsaev,
%``Light cone form of field dynamics in Anti-de Sitter space-time and AdS / CFT correspondence,''
 Nucl. Phys. B
\textbf{563} (1999), 295-348
%doi:10.1016/S0550-3213(99)00554-4
[arXiv:hep-th/9906217 [hep-th]].

\bibitem{Metsaev:2006}
R. R. Metsaev, {\it Nucl.Phys.} {\bf B759} (2006) 147–-201,
[hep-th/0512342].

%\cite{Metsaev:2007rn}
\bibitem{Metsaev:2007rn}
  R. R. Metsaev,
  %``Cubic interaction vertices for fermionic and bosonic arbitrary spin fields,''
{\it Nucl.Phys.} {\bf B859} (2012) 13
 % doi:10.1016/j.nuclphysb.2012.01.022
  [arXiv:0712.3526].
  %%CITATION = doi:10.1016/j.nuclphysb.2012.01.022;%%
  %93 citations counted in INSPIRE as of 16 Oct 2017

%\cite{Ivanovskiy:2023aay}
\bibitem{Ivanovskiy:2023aay}
V.~Ivanovskiy and D.~Ponomarev,
%``Light-cone formalism for a point particle in a higher-spin background,''
JHEP \textbf{09} (2023), 014
%doi:10.1007/JHEP09(2023)014
[arXiv:2306.13441 [hep-th]].


%\cite{Giombi:2012ms}
\bibitem{Giombi:2012ms}
  S.~Giombi and X.~Yin,
  %``The Higher Spin/Vector Model Duality,''
  J.\ Phys.\ A {\bf 46} (2013) 214003
  [arXiv:1208.4036 [hep-th]].


\bibitem{Vasiliev:1992av} M.~A.~Vasiliev, {\it Phys. Lett.}  B {\bf 285} (1992) 225.



%\cite{Didenko:2018fgx}
\bibitem{Didenko:2018fgx}
V.~Didenko, O.~Gelfond, A.~Korybut and M.~Vasiliev,
%``Homotopy Properties and Lower-Order Vertices in Higher-Spin Equations,''
J. Phys. A \textbf{51} (2018) no.46, 465202
%doi:10.1088/1751-8121/aae5e1
[arXiv:1807.00001 [hep-th]].





%\cite{Didenko:2019xzz}
\bibitem{Didenko:2019xzz}
V.~Didenko, O.~Gelfond, A.~Korybut and M.~Vasiliev,
%``Limiting Shifted Homotopy in Higher-Spin Theory and Spin-Locality,''
JHEP \textbf{12} (2019), 086
%doi:10.1007/JHEP12(2019)086
[arXiv:1909.04876 [hep-th]].

%\cite{Gelfond:2019tac}
\bibitem{Gelfond:2019tac}
O.~Gelfond and M.~Vasiliev,
%``Spin-Locality of Higher-Spin Theories and Star-Product Functional Classes,''
JHEP \textbf{03} (2020), 002
%doi:10.1007/JHEP03(2020)002
[arXiv:1910.00487 [hep-th]].


%\cite{Didenko:2020bxd}
\bibitem{Didenko:2020bxd}
V.~E.~Didenko, O.~A.~Gelfond, A.~V.~Korybut and M.~A.~Vasiliev,
%``Spin-locality of $\eta^{2}$ and $ {\overline{\eta}}^2 $ quartic higher-spin vertices,''
JHEP \textbf{12} (2020), 184
%doi:10.1007/JHEP12(2020)184
[arXiv:2009.02811 [hep-th]].

%\cite{Gelfond:2021two}
\bibitem{Gelfond:2021two}
O.~A.~Gelfond and A.~V.~Korybut,
%``Manifest form of the spin-local higher-spin vertex $\varUpsilon ^{\eta \eta }_{\omega CCC}$,''
Eur. Phys. J. C \textbf{81} (2021) no.7, 605
%doi:10.1140/epjc/s10052-021-09401-4
[arXiv:2101.01683 [hep-th]].


%\cite{Didenko:2022qga}
\bibitem{Didenko:2022qga}
V.~E.~Didenko,
%``On holomorphic sector of higher-spin theory,''
JHEP \textbf{10} (2022), 191
%doi:10.1007/JHEP10(2022)191
[arXiv:2209.01966 [hep-th]].

%\cite{Sharapov:2022nps}
\bibitem{Sharapov:2022nps}
A.~Sharapov, E.~Skvortsov, A.~Sukhanov and R.~Van Dongen,
%``More on Chiral Higher Spin Gravity and convex geometry,''
Nucl. Phys. B \textbf{990} (2023), 116152
%doi:10.1016/j.nuclphysb.2023.116152
[arXiv:2209.15441 [hep-th]].

%\cite{Vasiliev:2022med}
\bibitem{Vasiliev:2022med}
M.~A.~Vasiliev,
%``Projectively-compact spinor vertices and space-time spin-locality in higher-spin theory,''
Phys. Lett. B \textbf{834} (2022), 137401
%doi:10.1016/j.physletb.2022.137401
[arXiv:2208.02004 [hep-th]].

%\cite{Vasiliev:2016xui}
\bibitem{Vasiliev:2016xui}
M.~A.~Vasiliev,
%``Current Interactions and Holography from the 0-Form Sector of Nonlinear Higher-Spin Equations,''
JHEP \textbf{10}, 111 (2017)
%doi:10.1007/JHEP10(2017)111
[arXiv:1605.02662 [hep-th]].



%\cite{Vasiliev:2017cae}
%\bibitem{Vasiliev:2017cae}
%M.~A.~Vasiliev,
%``On the Local Frame in Nonlinear Higher-Spin Equations,''
%JHEP \textbf{01} (2018), 062
%doi:10.1007/JHEP01(2018)062
%[arXiv:1707.03735 [hep-th]].

%\cite{Gelfond:2017wrh}
\bibitem{Gelfond:2017wrh}
O.~A.~Gelfond and M.~A.~Vasiliev,
%``Current Interactions from the
%One-Form Sector of Nonlinear Higher-Spin Equations,''
Nucl. Phys. B
\textbf{931} (2018), 383-417
%doi:10.1016/j.nuclphysb.2018.04.017
[arXiv:1706.03718 [hep-th]].

%\cite{GV}
\bibitem{GV}
O.~A.~Gelfond and M.~A.~Vasiliev, unpublished.


%\cite{Vasiliev:1988sa}
\bibitem{Vasiliev:1988sa}
M.~A.~Vasiliev,
%``Consistent Equations for Interacting Massless Fields of All Spins in the First Order in Curvatures,''
Annals Phys. \textbf{190} (1989), 59-106.



%\cite{Tarusov:2022qpo}
\bibitem{Tarusov:2022qpo}
A.~A.~Tarusov, K.~A.~Ushakov and M.~A.~Vasiliev,
%``Shifted homotopy analysis of the linearized higher-spin equations in arbitrary higher-spin background,''
JHEP \textbf{03} (2023), 128
%doi:10.1007/JHEP03(2023)128
[arXiv:2212.01908 [hep-th]].


%\cite{DeFilippi:2019jqq}
\bibitem{DeFilippi:2019jqq}
D.~De Filippi, C.~Iazeolla and P.~Sundell,
%``Fronsdal fields from
%gauge functions in Vasiliev\textquoteright{}s higher spin
%gravity,''
JHEP \textbf{10} (2019), 215
%doi:10.1007/JHEP10(2019)215
[arXiv:1905.06325 [hep-th]].


%\cite{DeFilippi:2021xon}
\bibitem{DeFilippi:2021xon}
D.~De Filippi, C.~Iazeolla and P.~Sundell,
%``Metaplectic representation and ordering (in)dependence in Vasiliev\textquoteright{}s higher spin gravity,''
JHEP \textbf{07} (2022), 003
%doi:10.1007/JHEP07(2022)003
[arXiv:2111.09288 [hep-th]].

%\cite{Vasiliev:1999ba}
\bibitem{Vasiliev:1999ba}
  M.~A.~Vasiliev,
  %``Higher spin gauge theories: Star-product and AdS space,''
  arXiv:hep-th/9910096.

%\cite{Didenko:2022eso}
\bibitem{Didenko:2022eso}
V.~E.~Didenko and A.~V.~Korybut,
%``On z-dominance, shift symmetry and spin locality in higher-spin theory,''
JHEP \textbf{05} (2023), 133
%doi:10.1007/JHEP05(2023)133
[arXiv:2212.05006 [hep-th]].

%\cite{OG}
\bibitem{OG}
%\cite{Gelfond:2023fwe}
%\bibitem{Gelfond:2023fwe}
O.~A.~Gelfond,
%``Moderately non-local $\eta \bar\eta$ vertices in the $AdS_4$ higher-spin gauge theory,''
[arXiv:2308.16281 [hep-th]].


%\cite{Vasiliev:2003ev}
\bibitem{Vasiliev:2003ev}
M.~A.~Vasiliev,
%``Nonlinear equations for symmetric massless higher spin fields in (A)dS(d),''
Phys. Lett. B \textbf{567} (2003), 139-151
%doi:10.1016/S0370-2693(03)00872-4
[arXiv:hep-th/0304049 [hep-th]].

%\cite{Galperin:1984av}
\bibitem{Galperin:1984av}
A.~Galperin, E.~Ivanov, S.~Kalitsyn, V.~Ogievetsky and E.~Sokatchev,
%``Unconstrained N=2 Matter, Yang-Mills and Supergravity Theories in Harmonic Superspace,''
Class. Quant. Grav. \textbf{1} (1984), 469-498
[erratum: Class. Quant. Grav. \textbf{2} (1985), 127].
%doi:10.1088/0264-9381/1/5/004

%\cite{Buchbinder:2022svx}
\bibitem{Buchbinder:2022svx}
I.~Buchbinder, E.~Ivanov and N.~Zaigraev,
%``Unconstrained $\mathcal{N} = 2$ Higher-Spin Gauge Superfields and Their Hypermultiplet Couplings,''
Phys. Part. Nucl. Lett. \textbf{20} (2023) no.3, 300-305
%doi:10.1134/S1547477123030172
[arXiv:2211.09501 [hep-th]].

%\cite{Buchbinder:2022vra}
\bibitem{Buchbinder:2022vra}
I.~Buchbinder, E.~Ivanov and N.~Zaigraev,
%``$ \mathcal{N} $ = 2 higher spins: superfield equations of motion, the hypermultiplet %supercurrents, and the component structure,''
JHEP \textbf{03} (2023), 036
%doi:10.1007/JHEP03(2023)036
[arXiv:2212.14114 [hep-th]].

%\cite{Ivanov:2023uik}
\bibitem{Ivanov:2023uik}
E.~Ivanov,
%``Higher Spins in Harmonic Superspace,''
[arXiv:2306.10401 [hep-th]].


\end{thebibliography}
\end{document}